\documentclass[aps,prl,twocolumn,superscriptaddress]{revtex4-1}
\usepackage{graphicx}
\usepackage{soul}
\usepackage{color}
\usepackage{bm}        % for math
\usepackage{amsmath}
\usepackage{amssymb}   % for math
\usepackage{array}
\usepackage[caption=false]{subfig}
\usepackage[bookmarks=false,linkcolor=blue,urlcolor=blue,colorlinks,citecolor=blue]{hyperref}
\makeatletter

\begin{document}

\title{Superconductivity and non-Fermi liquid behavior near a nematic quantum critical point}

\author{Samuel Lederer}
\affiliation{Department of Physics, Massachusetts Institute of Technology, Cambridge, MA 02139, USA}
\author{Yoni Schattner}

\affiliation{Department of Condensed Matter Physics, The Weizmann Institute of Science, Rehovot, 76100, Israel}

\author{Erez Berg}
\affiliation{Department of Condensed Matter Physics, The Weizmann Institute of Science, Rehovot, 76100, Israel}
\author{Steven A. Kivelson}
\affiliation{Department of Physics, Stanford University, Stanford, CA 94305, USA}

\date{\today}

\begin{abstract}
Using determinantal quantum Monte Carlo, we compute the properties of a lattice model with spin $\frac 1 2$ itinerant electrons tuned through a quantum phase transition to an Ising nematic phase. The nematic  fluctuations induce  superconductivity with a  broad dome in the superconducting  $T_c$ enclosing the nematic quantum critical point.  For temperatures above $T_c$, we see strikingly non-Fermi liquid behavior, including a ``nodal - anti nodal dichotomy'' reminiscent of that seen in several transition metal oxides. {In addition, the critical fluctuations have a strong effect on the low frequency optical conductivity, resulting in behavior consistent with ``bad metal" phenomenology.} %SL and ``bad metal'' behavior of the conductivity.
\end{abstract}

% Optional adjustment to line up main text (after abstract) of first page with line numbers, when using both lineno and twocolumn options.
% You should only change this length when you've finalised the article contents.
%\verticaladjustment{-2pt}

\maketitle

Upon approach to a quantum critical point (QCP), the correlation length, $\xi$, associated with order parameter fluctuations diverges; consequently microscopic aspects of the physics %get
are averaged out and  certain   properties of the system  are universal.  Asymptotically close to criticality,  {\it exact} theoretical predictions concerning the scaling behavior of some measurable quantities are possible.  However, in solids, it is rarely possible to convincingly access asymptopia;   there are few %EBif any
experimentally documented cases in which a thermodynamic susceptibility
grows as a function of decreasing temperature, $T$, in proportion to a single power law $\chi \sim T^{-x}$ over significantly more than one decade of magnitude.  This is particularly true of metallic QCPs, where the  metallic critical point may be preempted by the occurrence of a superconducting dome, a fluctuation driven first order transition, or some other catastrophe.

However, there is a looser sense in which a QCP
can serve as an organizing principle for understanding properties of solids over a
  range of  parameters:  In the ``neighborhood'' of a QCP, where $\chi$ is large (in natural units) and $\xi$ is more than a
  few lattice constants, it is reasonable to conjecture that quantum critical fluctuations play a significant role in determining the properties of the material and that, at least on a qualitative level, those properties may be robust ({\it i.e.} not strongly dependent on microscopic details), even if they are not universal.

 \begin{figure}
\includegraphics[clip=true,trim= 30 220 10 220, width=\columnwidth]{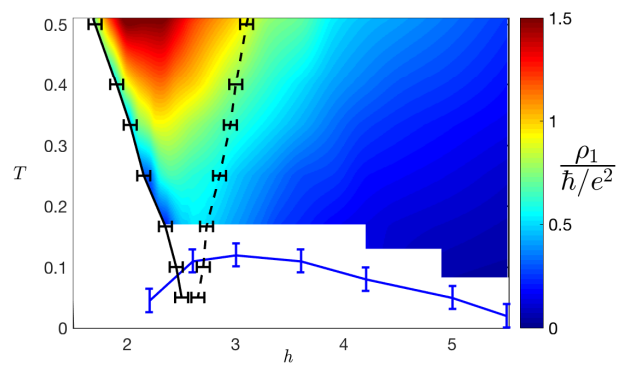}
\caption{Phase diagram
as a function of the transverse field $h$ and temperature $T${, both measured in units of the fermion hopping $t$}. The solid and dashed black lines indicate, respectively, the boundary of the nematic phase ($T_{nem}$) and a crossover  ($T^\star_{nem}${, defined shortly before \eqref{Gofbeta}}) marking the edge of the quantum critical fan. The superconducting $T_c$ is indicated by the blue solid line. The color scale shows  {a proxy for} the DC resistivity in units of $\hbar/e^2$, defined from the fit in \eqref{eq:lambda_fit_mats}. %SLinferred from the DQMC data as described in the text.
Here, $\alpha = 1.5$, $V=0.5 t$, $\mu = t$.  Error bars reflect finite size errors, as described in the Supplement.}
\label{fig:phase_diagram}
\end{figure}

With this in mind, we have carried out extensive numerical ``experiments'' on a simple 2D lattice model of itinerant electrons coupled to an Ising-like ``nematic'' order parameter field, Eq.~\ref{eq:HFHB}.  By  varying a  parameter in the Hamiltonian, $h$, the system can be tuned through a quantum or thermal transition from a disordered (symmetric) phase to a nematic phase that spontaneously breaks the lattice symmetry from $C_4$ to $C_2$.  %SL The continuum version of this model has
{Related models of nematic quantum criticality have} been studied extensively
\cite{Oganesyan2001,metzner-2003,kim2004pairing,Dellanna2006,Lawler2006,Metlitski2010,mross2010controlled,Maslov2010,fitzpatrick1,Dalidovich2013,metlitski2015cooper,Holder2015,Thier2011,Mandal2015,lederer2015enhancement,Eberlein2016,Mandal2016,Mandal2016b,Patel2016,Punk2016,Hartnoll2014,Drukier2012,Mahajan2013,Fitzpatrick2014,Raghu2015,Holder2015b,Meszena2016}
using {various analytic methods}, %SLfield theoretic methods; %EBIn part, this model is interesting because
%SL  it can
and can also be studied with minus-sign-free determinantal quantum Monte Carlo (DQMC)~\cite{Blankenbecler1981,scalettar-1989,AssaadBook}.
 %EB with  results that may  be representative of a broad class of metallic QCPs.
 Moreover, the model is particularly topical, as there is good evidence that a nematic QCP underlies the superconducting dome in many (possibly all) of the Fe-based superconductors~\cite{song2011direct,Chu2012, Gallais2013,  zhou2013quantum, Bohmer2014, Blumberg2016,  kuo2016ubiquitous} and possibly even the cuprate high temperature superconductors~\cite{hinkov-2008,taillefer-nematic-2009,lawler2010intra}. This prospect has recently been explored using Monte Carlo methods in \cite{Schattner2016, Li2015, Dumitrescu2016, Xu2016}.

In a previous study~\cite{Schattner2016} of %SL same
this
 model, we focused on the critical scaling at the putative metallic QCP
 {with a moderate
 dimensionless coupling
  between the itinerant electrons and the nematic fluctuations
 $\alpha=1/2$. }%SL We took the  coupling, $\alpha$, between the itinerant electrons and the nematic fluctuations, %to be reasonably small, %SL$\alpha\leq 1/2$;
 We found that, though superconducting fluctuations are enhanced when the tuning parameter $h$ is close to the quantum critical value, $h_{c}$, the superconducting $T_c$ (if any) is below the accessible range of temperatures.
{We documented a possible mild breakdown of Fermi liquid behavior, and identified a broad range of small $T$ and $|h-h_{c}|$ in which some correlation functions are well approximated by simple scaling functions.}
However, other correlations with the same symmetries
do not exhibit the same scaling behavior.
This implies that the regime we
 accessed is
 far enough from criticality that corrections to scaling
 are significant, or that the scaling behavior we see does not reflect the properties of a metallic QCP at all.

In the present paper, we have focused on the
properties of the model in the critical ``neighborhood,'' and
  have taken larger values of $\alpha = 1 - 1.5$ so that all the energy scales are enhanced, making them easier to document in numerical experiments.  Our %principle
  principal findings are:

{ 1) } As shown in
Fig.~\ref{fig:phase_diagram}, there is a broad superconducting dome with its $T_c$ maximum roughly coincident with the value $h=h_c$ at which the nematic transition temperature, $T_{nem}\to 0$.  {The maximum $T_c$ is found to be about $0.03 E_F$, where $E_F$ is the Fermi energy.} 

{ 2)}  Fig.~\ref{fig:heatmap} shows the electron spectral function, $A(\vec k,\omega)$, integrated over a range of energies of order $T$ about the Fermi energy, as defined in Eq. \ref{Gofbeta}.
The nematic fluctuations are
exceedingly effective in destroying quasiparticles,
and indeed produce a striking ``nodal - anti nodal'' dichotomy in which some remnant of the Fermi surface is still visible along the zone diagonal.  This is associated with the existence of ``cold-spots''~\cite{Metzner2003,Kyungmin2016} on the Fermi surface
%, which
that are required by symmetry.
%{ the $x^2-y^2$ symmetry of the nematic order parameter.}
Away from the cold-spots, the imaginary part of the (Matsubara) self energy, shown in Fig. 3, is dramatically unlike that of a Fermi liquid throughout the quantum critical regime.

{3)} { Transport properties are also strongly affected in the quantum critical regime above $T_c$. Since analytic continuation of imaginary time data to real time is numerically ill-posed, %SL we cannot access the DC limit,
we have carried out two different procedures to obtain {\it proxies} for the resistivity. These proxies, plotted in Fig.~\ref{fig:rho_T},  track the DC resistivity under certain assumptions described below. 
The first proxy, $\rho_1$, is derived from a simple two-component fit to  the DQMC data (shown in Fig.~\ref{fig:lambda}) while the second, $\rho_2$, defined in Eq. \ref{proxy}, is extracted directly from the long (imaginary) time behavior of the current-current correlation function.  The results obtained by the two methods are qualitatively similar.
At $h=h_c$ both proxies are roughly linearly increasing functions of $T$. For the larger value of the coupling constant, the proxies exceed the  quantum of resistance, $\rho_q=\hbar/e^2$, at high temperatures (but still much below $E_F$). Outside the quantum critical fan,
%the resistivity
 the proxies are substantially smaller than $\rho_q$ for all $T$.}
 %temperatures %EBand has a weaker temperature dependence (consistent with $\rho \sim T^2$).
(See also Fig. 1.)
\section*{The Model}
Our model is defined on a two-dimensional square lattice, where every site
has a single Wannier orbital.
 Each link has a pseudospin-$1/2$ degree of freedom that couples to the fermion bond-density. The system is described by the
  Hamiltonian %:
\begin{eqnarray}
H &=& H_f + H_b + H_\mathrm{int}, \nonumber \\
H_f &=& -t\sum_{\langle i,j \rangle, \sigma} c^\dagger_{i\sigma} c^{\vphantom{\dagger}}_{j\sigma} - \mu \sum_{i, \sigma} c^\dagger_{i\sigma} c^{\vphantom{\dagger}}_{i\sigma}, \nonumber \\
H_b &=& V\sum_{\langle \langle i,j\rangle;\langle k,l \rangle \rangle} \tau^z_{i,j} \tau^z_{k,l} -h \sum_{\langle i,j\rangle } \tau^x_{i,j}, \nonumber \\
H_\mathrm{int} &=& \alpha t \sum_{\langle i,j\rangle, \sigma} \tau^z_{i,j} c^\dagger_{i\sigma} c^{\vphantom{\dagger}}_{j\sigma},
\label{eq:HFHB}
\end{eqnarray}
where $c^\dagger_{j\sigma}$ creates a fermion on site $j$ with spin $\sigma=\uparrow,\downarrow$, $\langle
i, j \rangle$ denotes a pair of nearest-neighbor sites, $t$ and $\mu$ are the hopping strength and chemical potential, respectively, $\tau^a_{i,j}$ ($a = x,y,z$) denote
pseudospin 1/2 operators that live on the bond connecting the neighboring sites $i$ and $j$, $V>0$ is the
 Ising interaction between nearest-neighbor
 pseudospins (here, $\langle \langle i,j\rangle;\langle k,l \rangle \rangle$ denotes a pair of nearest-neighbor bonds), $h$ is the strength of a transverse field that acts on the pseudospins, and $\alpha$ is the dimensionless coupling strength between the pseudospin and the fermion bond density. { In the ordered phase where $V$ dominates, $\tau^z_{i,j}$ adopts a staggered configuration, taking different values on horizontal and vertical bonds, thereby generating nematic order.}

\section*{DQMC Results}
A typical phase diagram is shown in Fig.~\ref{fig:phase_diagram}, for
$\alpha=1.5,V=0.5 t, \mu= t$.
~\footnote{A value of $\alpha>1$ is problematic microscopically, since the effective hopping matrix element along one direction changes sign deep in the ordered phase.
%EBWe have limited our analysis to the region of the $h-T$ plane in which both effective hoppings are of the same sign.
%Nevertheless, treating
However, we view Eq.~(\ref{eq:HFHB}) as an effective model designed to give a nematic QCP, %we
and so do not restrict the value of $\alpha$.}
{(%hereafter
Hereafter, we use units in which  $t=\hbar=e^2=1$).} %.
In addition to the
  nematic and symmetric phases, there is a ``dome" of superconductivity with maximum critical temperature near $h_c$, as anticipated~\cite{kim2004pairing,Maier2014,metlitski2015cooper,lederer2015enhancement}. The   pair wave-function in the superconducting state has
  spin singlet s-wave symmetry in the %SLisotropic
{symmetric} phase and mixed s and d-wave symmetry in the nematic phase.

The boundary of the nematic phase $T_{nem}$ and the crossover temperature
$T_{nem}^\star$ are both
derived from an analysis of the thermodynamic nematic susceptibility
$\chi (h,T)  \equiv
\frac{1}{L^2} \sum_{i,j} \int_0^\beta d\tau \langle N_{i}(\tau) N_{j}(0) \rangle$,
where the nematic order parameter is defined as $N_i = \sum_j \eta_{ij} \tau^z_{ij}$, where $\eta$ is a d-wave form factor: $\eta_{ij}=1/4$ for $\mathbf{r}_{ij} = \pm \hat{\mathbf{x}}$, $\eta_{ij}=-1/4$ for $\mathbf{r}_{ij} = \pm \hat{\mathbf{y}}$, and $\eta_{ij}=0$ otherwise. $T_{nem}$ is determined using finite size scaling appropriate to a two-dimensional classical Ising transition, %EB\cite{finitesizescaling},
while
$T^\star_{nem}(h)$
is defined
implicitly according to $\chi(h,T_{nem}^*)=\frac 1 2 \chi(h_c,T_{nem}^\star)$. The superconducting critical temperature $T_c$ is determined by analysis of the superfluid stiffness~\cite{Scalapino1993,Paiva2004}, and can also be estimated by other %EB thermodynamic and spectroscopic
methods, yielding similar results~\cite{SOM}. %EBFurther details %on the determination of phase boundaries

{ The presence of superconductivity limits the region in which any scaling behavior of nematic fluctuations can be identified. That said, for temperatures well above $T_c$, the thermodynamic nematic susceptibility near $h_c$ is similar in structure to that reported in \cite{Schattner2016}. Nematic fluctuations at nonzero frequency have somewhat different structure than those previously reported, and have a reduced dependence on momentum.}

Turning
to single-particle properties, we examine
\cite{Trivedi1995}
\begin{equation}
{\cal G}(\vec k)\equiv 2 \widetilde G\left(\vec k,\tau=\frac{\beta}{2}\right)=\int d\omega\frac{A(\vec k, \omega)}{\cosh\left[{\beta\omega}/{2}\right]}
\label{Gofbeta}
\end{equation}
where $\widetilde G$ is the imaginary time fermion Green function and $A$ is the real frequency spectral function. Roughly, $\cal G$ measures spectral weight within an energy range of order $T$ of the Fermi level, so that a
 sharp peak in the momentum dependence of $\cal G$ indicates an underlying Fermi surface. In a Fermi liquid, ${\cal G}(\vec k)$
  is
 peaked at the Fermi surface
 with a peak amplitude that approaches the quasiparticle residue $Z_{\vec k}$ as $T\to 0$.

 \begin{figure}
\includegraphics[clip=true,trim= 135 195 47 225, width=0.9 \columnwidth]{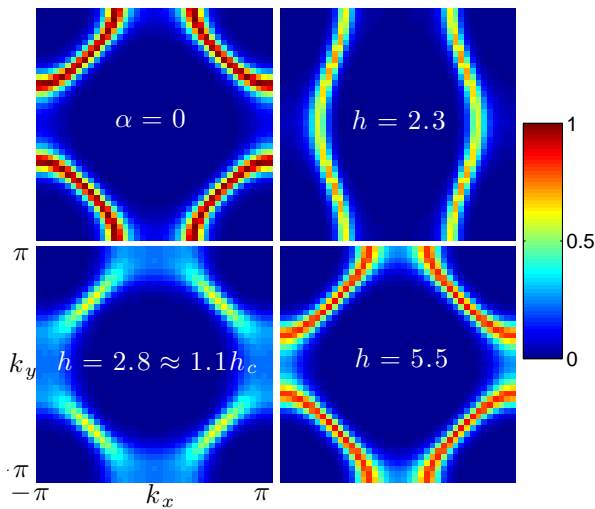}
\caption{The low frequency fermionic spectral weight ${\cal G}(\vec k)$ [Eq. (\ref{Gofbeta})] at temperature $T=0.17$, shown for the noninteracting band structure ($\alpha=0$),
and for $\alpha=1.5$ for several values of $h$.
For $h=2.3$, a small $C_4$ symmetry-breaking field has been applied in the simulation. Data are for a $20\times20$ system with various combinations of periodic and antiperiodic boundary conditions. Other parameters are $V=0.5 $ and, $\mu = 1.0$.}
\label{fig:heatmap}
\end{figure}

Fig.~\ref{fig:heatmap} shows color plots of $\cal G$ at fixed temperature $T=0.17$
 for the non-interacting case ($\alpha=0$) and for $\alpha=1.5$ at a variety of values of $h$.
 For $h=5.5$, deep in the symmetric phase, there is a clear Fermi surface similar to that of the non-interacting problem. For $h=2.3$, in the ordered phase, there is also a clear Fermi surface consisting of open sheets, reflecting a substantial nematic distortion. For $h=2.8$, near $h_c$, a Fermi surface appears well-defined near the
 cold spots along the zone diagonal, but is increasingly ill-defined away from these high-symmetry points. The higher degree of coherence along the zone diagonal is expected by symmetry: %since nematic order breaks diagonal reflection symmetry,
 long-wavelength nematic fluctuations cannot couple to fermions with momentum along the zone diagonal.

 \begin{figure}
\includegraphics[clip=true,trim= 70 200 20 210, width=\columnwidth]{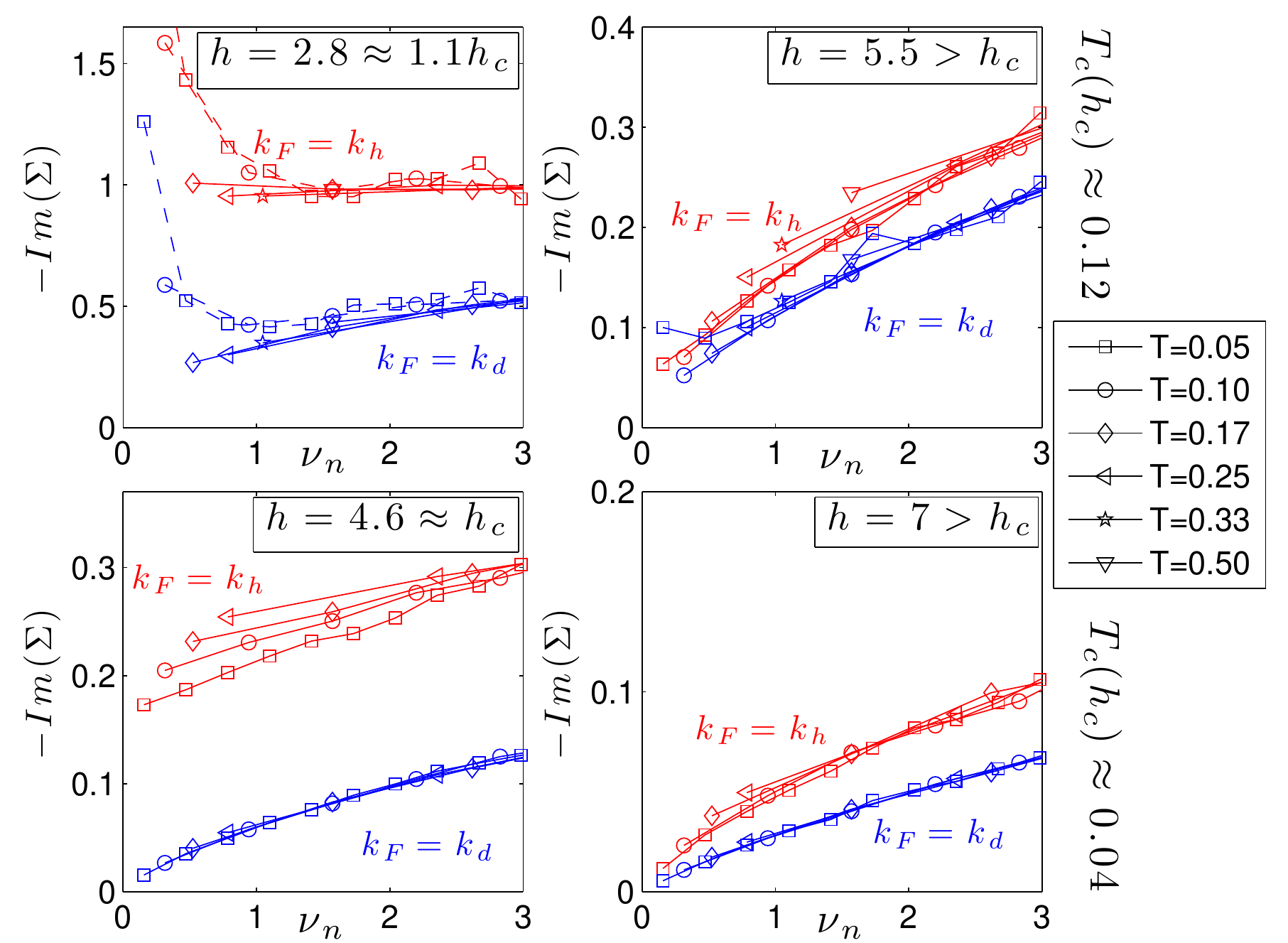}
\caption{The imaginary part of the fermion self-energy for two values of parameters: $\alpha=1.5,V=0.5,\mu=1.0$ (top row), and $\alpha=1.0, V=1.0, \mu=0.5$ (bottom row), for various temperatures, and %SL two choices of
%{\color {red}
with the nominal Fermi momenta  $\vec k_d$ and $\vec k_h$ along the $(0,0)-(\pi,\pi)$ and $(0,\pi)-(\pi,\pi)$ %SL diagonal and horizontal
  directions, respectively. %}%SLhigh-symmetry directions.
Data are shown for a $20\times 20$ system both near $h_c$ (left column) and far in the symmetric phase (right column). In the upper left panel, data points below $T_c$ are connected by dashed lines. }
\label{fig:Sigma}
\end{figure}

A complementary probe of single fermion properties is the self energy, extracted from the
Green function $G(\vec k,\nu_n)$ and the non-interacting Green function $G_0(\vec k,\nu_n)$ according to $\Sigma(\vec k, \nu_n)\equiv G_0^{-1}(\vec k, \nu_n)-G^{-1}(\vec k, \nu_n)$.  (Henceforth, $\nu_n=2\pi T(n+1/2)$ and $\omega_n=2\pi Tn$ will designate the fermionic and bosonic Matsubara frequencies.)
 In a Fermi liquid at asymptotically low temperature, $-\mathrm{Im}[\Sigma(\vec k_F,\nu_n)]=\gamma_{\vec k}+(1/Z_{\vec k}-1)\nu_n +{\cal O}(\nu_n^2)$ for $\nu_n>0$, where $\gamma_{\vec k}(T) \ll T$ is
 the inelastic scattering rate.
 More generally, $\gamma_{\vec k}$ is obtained by extrapolating  $-\mathrm{Im}[\Sigma]$ to zero frequency.
 In Fig.~\ref{fig:Sigma} {we plot $-\mathrm{Im}[\Sigma(\vec k,\nu_n)]$ vs. $\nu_n$, %either
 both close to $h_c$ %or
 and deep in the disordered phase.} %EB we plot $-\mathrm{Im}[\Sigma(\vec k,\nu_n)]$ vs. $\nu_n$ for parameters $\alpha=1.0, V=1.0,\mu=-0.5$, at $h=4.6 \approx h_c$, and at $h=7.0$, deep in the ordered phase.
 We show data for a variety of temperatures
 for $\vec k$ at the nominal Fermi momenta $\vec k_d$ and $\vec k_h$ along the %{\color{red}
 $(0,0)-(\pi,\pi)$ and $(0,\pi)-(\pi,\pi)$ %SL diagonal and horizontal
 directions, respectively. %}

{In the disordered phase,} the frequency and temperature dependences of $\mathrm{Im}[\Sigma]$ at both $\vec k_d$ and $\vec k_h$
are consistent with Fermi liquid theory -- the extrapolated $\nu_n\to 0$ intercept ($\gamma_{\vec k}$) is well below $T$
and the slope is
finite (corresponding to $1>Z_{\vec k}>0$) and hardly $T$ dependent.
Even for $h\approx h_c$, Fermi liquid theory is loosely consistent with the data at $\vec  k_d$, but
not remotely so at $\vec k_h$ where
$\gamma_{\vec k_h}$ exceeds $T$, and appears not to
vanish in the $T\to 0$ limit.
In the quantum critical regime and above $T_c$, quasiparticles far from the cold spots are not even marginally well-defined. %{
(The upturn of $\mathrm{Im}[\Sigma]$ at low frequency, visible especially for $\alpha=1.5$, is associated with the onset of a superconducting gap.)
% }

The intervention of superconductivity complicates any analysis of the putative low temperature Fermi liquid properties as $h \to h_c$.
However, to obtain a rough sense of trends, one can estimate the dispersion of the quasiparticle-like features as a function of $h$ and $T$ at different parts of the Fermi surface (see Supplementary Information for details). We see a tendency for the dispersion to become substantially flatter as $h\to h_c$ ({\it i.e.} a large increase in the ``effective mass''), though any such renormalization is much weaker or non-existent at the cold spots on the Fermi surface. { The electronic spectral function $A(\vec{k}, \omega)$, calculated from $G(\vec{k}, \nu_n)$ using the maximum entropy method, is consistent with such behavior (see Supplementary Information). Near the cold spots $A(\vec{k}, \omega)$ has a well-defined dispersive peak, while in the hot regions at $h\approx h_c$ there are only broad features without a clear dispersion. Below the superconducting $T_c$, $A(\vec{k}, \omega)$ clearly displays a superconducting gap in both the cold spots and the hot regions (with a larger gap in the hot regions).}

%EB However,
%to obtain a rough sense of trends, we have computed the $\vec k$ dependence of the appropriate moment of $G$ (defined in the Supplemental Material and shown in Fig. ***) to  obtain an estimate of a quasiparticle-like dispersion;  we see a clear tendency for the dispersion to become substantially flatter as $h\to h_c$ ({\it i.e.} a large increase in the ``effective mass''), while any such renomralization is much weaker or non-existent at the cold spots on the Fermi surface.  }
%The strong violation of Fermi liquid heory in single particle properties near the QCP {motivates us to look for non-Fermi liquid behavior also in the electrical conductivity.}
%SL The electrical conductivity near the QCP is also inconsistent with Fermi liquid theory.

{ The breakdown of Fermi liquid theory seen in the fermion Green function suggests that transport properties may also be strongly altered near the QCP. %EB, but here our analysis is necessarily more indirect.
%SLThe quantity of greatest interest is %EBof course
One quantity of great interest is the DC conductivity, but the DC limit of transport is particularly difficult to access using analytic continuation of imaginary time data. %EBinaccessible due to the ill-posed nature of numerical analytic continuation.
The analysis we carry out below yields %EBwell-defined statements
information about the optical conductivity at frequencies of order the temperature, but any statements about the DC conductivity rest on additional, nontrivial assumptions.}

\begin{figure}
\includegraphics[clip=true,trim= 20 60 50 90, width=\columnwidth]{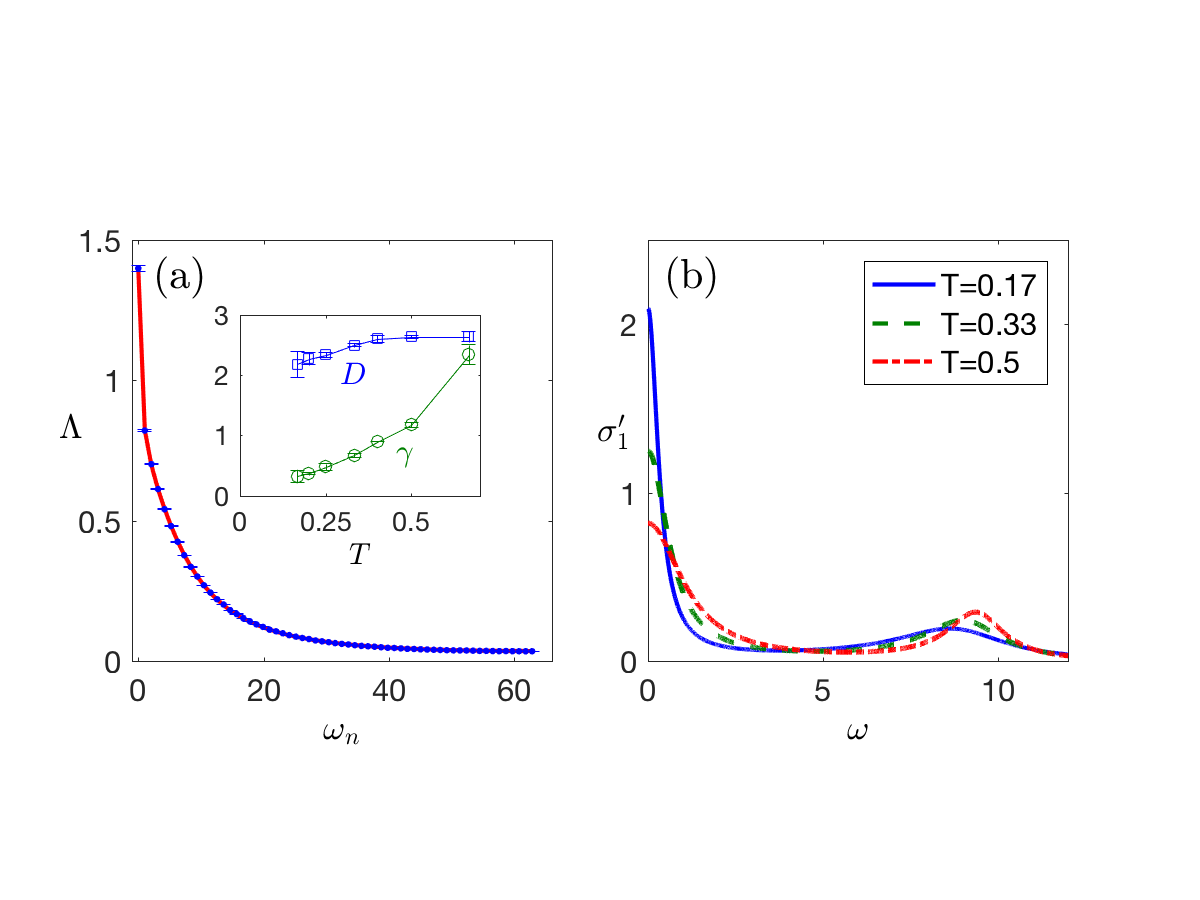}
\caption{Left: the current-current correlator for a $16\times 16$ system with $h=2.6\approx h_c,T=0.17$, for $\alpha=1.5, V=0.5, \mu = 1$. Error bars are comparable to the symbol size. The solid red line is a least-squares fit to two components of the form of \eqref{eq:lambda_fit_mats}. Right: the corresponding real part of the optical conductivity for parameters on the left, as well as two higher temperatures, showing how the Drude-like peak sharpens on cooling. The temperature dependence of the half-width at half-maximum $\Gamma$ of the Drude-like peak, as well as its weight $D$, are shown in the inset to the left panel, with error bars estimated as described in the supplement.}
\label{fig:lambda}
\end{figure}

We have measured the imaginary time ordered current-current correlator $\widetilde \Lambda_{ii}(\tau)\equiv \langle {\cal T} J_i(\tau) J_i(0)\rangle$, where $J_i$ is the uniform current operator in direction $i=x$ or $y$.
(We will henceforth leave the directional indices implicit.)
 $\Lambda(\omega_n)$, the Fourier transform of $\widetilde\Lambda(\tau)$, is shown in Fig.~\ref{fig:lambda}a for $\alpha=1.5,V=0.5,\mu=1,h\approx h_c$,  \& $T=0.17\approx 1.5 T_c$. In a non-superconducting state, $\Lambda(\omega_n)$ is related to the real part of the optical conductivity $\sigma'(\omega)$ by
\begin{align}
\Lambda(\omega_n)=&\int\frac{d\omega}{\pi} \frac{\omega^2\sigma' (\omega)}{\omega^2+\omega_n^2}.
\end{align}
A clear feature, present throughout the
non-superconducting portions of the phase diagram,
is a substantial jump in $\Lambda(\omega_n)$ between the zeroth and first Matsubara frequency. This is evidence of a Drude-like
 component of
 $\sigma^\prime(\omega)$
 peaked at low frequencies,
 with a width less than or comparable to $T$.  The slow decrease of $\Lambda(\omega_n)$ for $n>1$ is indicative of
 an additional broad
 feature with optical weight spread over a range of frequencies large compared to $T$.

We have performed a simple analytic continuation of our data via a least squares fit. The fitting function is a sum of two terms \footnote{Eq. \eqref{eq:lambda_fit_mats} must be modified to account for the discretization of imaginary time, as discussed in the Supplementary Material.}
%SL Given these observations, we have performed a least squares fit to our data as the sum of two terms,
\begin{eqnarray}
 \Lambda_{\text{fit}}(\omega_n)=\sum_{j=1}^2 \frac{A_j}{\omega_n^2+\gamma_j|\omega_n|+\Omega_j^2}.
\label{eq:lambda_fit_mats}
\end{eqnarray}
$\Lambda_{\text{fit}}$ can then be analytically continued to give
\begin{align}
\sigma^\prime(\omega)=&\sum_{j=1}^2\frac{A_j\ \gamma_j}{(\Omega_j^2-\omega^2)^2+ \gamma_j^2 \omega^2}.
\label{eq:lambda_fit_real}
\end{align}
As illustrated in Fig.~\ref{fig:lambda}a, the fit agrees with the data within {a few percent}.
The corresponding optical conductivity
is shown in Fig.~\ref{fig:lambda}b for a variety of temperatures above $T_c$; it consists of a Drude-like component with its maximum at $\omega=0$ (i.e. ${\gamma_1 > \sqrt{2}\Omega_1}$)
  that broadens with increasing temperature,
and a broad, largely temperature-independent background with a maximum at ${\omega=\sqrt{|\Omega_2|^2 - |\gamma_2|^2/2}>0}$. {The zero frequency limit of this fitted conductivity yields a proxy $\rho_1$ for the DC resistivity.}

Though physically plausible and in agreement with our data, the { fitting}
analysis is not unique -- analytic continuation of numerical data is a famously ill-conditioned problem~\cite{gubernatis1991quantum}. { As one check on our results, we have performed the analytic continuation using standard maximum entropy methods; the results,  as shown in the Supplementary Material, are very similar to those obtained above. On the other hand}{, as also shown in the Supplementary Material, the quality of the fit is similar if we mandate a third component with width far less than the temperature, which would of course drastically alter the DC conductivity. %EB, but would not alter the conductivity at frequencies $\sim T$.
Such a narrow peak may arise %SL for example, if umklapp processes are ineffective, and as a result
if there is an emergent nearly-conserved momentum\cite{Maslov2011,Mahajan2013}.} %SL in the theory.

{ Analysis of the current-current correlator in the time domain yields additional information. %EBAs described in the supplement,
The value and the derivatives of $\widetilde\Lambda(\tau)$ near $\tau=\beta/2$ contain information about the moments of the low frequency part of the optical conductivity:
\begin{equation}
\big[\partial^{2m}_\tau{\widetilde\Lambda} \big]_{\tau=\beta/2} = \int \frac{d\omega}{2\pi} \frac{\omega^{2m+1} \sigma'(\omega)}{\sinh\left(\frac{\beta \omega}{2}\right)}.
\end{equation}
The first two such %SL  low frequency
moments obtained from our QMC simulations are shown in the Supplementary Material. (Interestingly, these moments can also be straightforwardly computed from empirical data, enabling { direct} comparison with experiment.)

The two lowest order moments can be combined into a quantity with units of resistivity according to
\begin{equation}
\rho_2\equiv
\big[\partial^2_\tau{\widetilde\Lambda}\big/(2\pi\widetilde\Lambda^2)\big|_{\tau=\beta/2}.
\label{proxy}
\end{equation}
This quantity tracks the DC resistivity at low temperatures whenever the low frequency ($\omega \lesssim T$) conductivity can be described by a single Drude-like component which either has Lorentzian shape or a width of order $T$. This is a parsimonious  { (although  not unassailable)} assumption and consistent with our data.
%, but, as previously stated, we cannot rule out a more complicated structure at low frequencies. %EBmight arise due to emergent momentum conservation [NEED CITATION].
With caveats in place, we now describe the behavior of the two resistivity proxies $\rho_{1,2}$ defined above.}
%SLThus, to test the robustness of our conclusions, especially concerning the DC resistivity $\rho$,
%we additionally define a resistivity proxy $%\widetilde
%\rho_{\rm pr}$
%{which does not require analytic continuation:} \begin{equation}
%%\widetilde
%\rho_{\rm pr}\equiv
%\big[\partial^2_\tau{\widetilde\Lambda}\big/(2\pi\widetilde\Lambda^2)\big|_{\tau=\beta/2}.
%\label{proxy}
%\end{equation}
%As shown in the supplement, $%\widetilde
%\rho_{\rm pr}$ is
% proportional to the DC resistivity %at low temperature
% in a variety of
% circumstances.

\begin{figure}
\includegraphics[clip=true,trim= 80 190 50 210, width=\columnwidth]{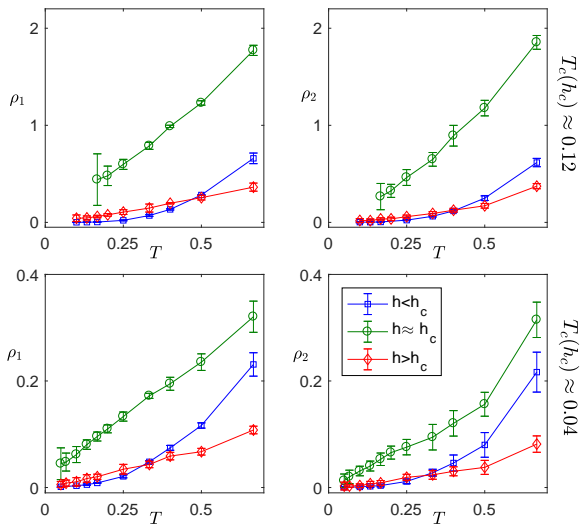}
\caption{The temperature dependence of the resistivity {proxies} (in units of $\hbar/e^2$) for parameters $\alpha=1.5,V=0.5, \mu=1$ (top row, for $h=2.0<h_c$, $h=2.6\approx h_c$ and $h=5.5>h_c$) and $\alpha=1.0, V=1.0,\mu=0.5$ (bottom row, for $h=4.0<h_c$, $h=4.6\approx h_c$ and $h=7.0>h_c$). For $h<h_c$, a small symmetry breaking field has been applied, and the smaller component of the resistivity { proxy} is shown. Values shown are for the largest system size simulated ($L=16$ or $18$ depending on temperature), with error bars estimated as described in the Supplementary Material.}
\label{fig:rho_T}
\end{figure}

%$\rho_2(h,T)$
{ $\rho_1(h,T)$} is represented in the coloring of the symmetric metallic region of the phase diagram in Fig.
 1. It exhibits a non-monotonic dependence on $h$, with a maximum near $h_c$. The temperature dependence of $\rho_1$ and $%\widetilde
  \rho_2$ are shown in Fig.~\ref{fig:rho_T} in the ordered and disordered phases as well as at $h_c$, for both $\alpha=1.5,V=0.5,\mu=1.0$ and $\alpha=1.0,V=1.0,\mu=0.5$. $\rho_1$ and $%\widetilde
  \rho_2$ are qualitatively similar over a wide range of temperatures. Both
 {are significantly higher} at $h\approx h_c$ than deep in the ordered and disordered phases.
 In the ordered phase, the data are roughly consistent with the $T^2$ temperature dependence expected of a Fermi liquid. In the disordered phase, the temperature dependence in the range of $T>T_c$ can be fit to a linear function of $T$ with small slope and a slightly negative extrapolated value at $T\to 0$.
\footnote{Recalling that small resistivities correspond to a sharp Drude-like feature in the resistivity -- precisely the sort of feature that is most difficult to capture reliably from imaginary time data -- we have not attempted a serious analysis of the apparently non-Fermi liquid character of this last observation.}

At $h\approx h_c$, the behavior depends somewhat on parameters. For $\alpha=1.0,V=1.0,\mu=0.5$, there is an apparent $T-linear$ behavior over about a decade of temperature. For $\alpha=1.5,V=0.5,\mu=1.0$, the high $T_c$ leaves an insufficient dynamical range to establish a  clear power law temperature dependence, but both $\rho_1$ and $\rho_2$ exceed the Ioffe-Regel limit of $\hbar/e^2$ at a temperature of approximately $3T_c$.
Subject always to the uncertainties in analytic continuation, the behavior of our model near $h_c$ is
strikingly reminiscent of the ``bad metal" phenomenology seen in many correlated materials~\cite{emery-1995,hussey-2004,hartnoll-2015}.

%To test the robustness of these results, w
We have performed additional simulations at lower fermionic densities, with results summarized in the supplementary material.
 Much of the phenomenology %is similar to the high densities described above
appears to be robust: Close to $h_c$, the imaginary part of the fermionic self-energy at $\vec k_h$ %seems to approach
approaches a constant %near $h_c$,
and the resistivity is of order of the quantum of resistance.  % close to $h_c$.
However, the temperature dependence of the resistivity is not linear. % close to $h_c$.
 Also, for certain values of the couplings, we find evidence that the nematic transition becomes weakly first order at low temperatures. %We defer a systematic study of the density dependence of the resistivity to a future publication.

\section*{Discussion}
 We have studied the vicinity of a nematic QCP in a simple lattice model of a metal.
 The QCP is masked by a dome-shaped superconducting phase. The normal-state quantum critical regime does not exhibit clear scaling behavior; however, it displays strong anomalies that we associate with the approach to the QCP. In particular, the fermion self-energy is strikingly non-Fermi liquid like over much of the Fermi surface.
{The optical conductivity at frequencies $\lesssim T$ is also strongly affected by the critical fluctuations. Assuming a simple form of $\sigma(\omega)$, we find that the DC resistivity is anomalously large (exceeding the Ioffe-Regel limit for  $\alpha > 1$) and nearly linear in temperature.}

While our model
 does not accurately describe the microscopics of any specific material, and ignores physical effects that may be important~\cite{Carlson2006,Nie2014,Karahasanovic2016,Paul2016},
 it is plausible that the qualitative behavior proximate to the QCP is
 relatively  insensitive to microscopic details. Our results bear striking similarities to the behavior seen in certain high temperature superconductors: in several iron-based superconductors, the resistivity
 is anomalously large and $T$-linear near the putative nematic QCP~\cite{Chu2012,matsuda2013}, and the fermionic spectral properties of our model in the critical regime are reminiscent of the ``nodal-antinodal dichotomy'' reported in the ``strange metal'' regime of certain cuprates~\cite{valla-2000,zhou-2004}.

\begin{acknowledgments}
{\it Acknowledgements.--} 
S.~L. and Y.~S. have contributed equally to this work. The authors acknowledge fruitful discussions with Andrey Chubukov, Snir Gazit, Sean Hartnoll, Mohit Randeria, Subir Sachdev, Boris Spivak, and Yochai Werman. S.~L. was supported by a Gordon and Betty Moore Post Doctoral Fellowship at MIT.
S.~A.~K. was supported in part by NSF grant DMR 1265593 at Stanford.
Y.~S. and E.~B. were supported by the Israel Science Foundation under
Grant No. 1291/12, by the US-Israel BSF under Grant No.
2014209, and by a Marie Curie reintegration grant. E. B. was
supported by an Alon fellowship.
This work was performed in part at the Aspen Center for Physics, which is supported by National Science Foundation grant PHY-1066293.
\end{acknowledgments}

\bibliography{fulldraft}
%merlin.mbs apsrev4-1.bst 2010-07-25 4.21a (PWD, AO, DPC) hacked
%Control: key (0)
%Control: author (8) initials jnrlst
%Control: editor formatted (1) identically to author
%Control: production of article title (-1) disabled
%Control: page (0) single
%Control: year (1) truncated
%Control: production of eprint (0) enabled
%

%\end{document}
\clearpage

\widetext
\begin{center}
\textbf{\large Supplementary Information for: \\ ``Superconductivity and bad metal behavior near a nematic quantum critical point'' }
\end{center}
\setcounter{equation}{0}
\setcounter{figure}{0}
\setcounter{table}{0}
\setcounter{page}{1}
\makeatletter

\renewcommand{\theequation}{S\arabic{equation}}
\renewcommand{\thesection}{S-\Roman{section}}
\renewcommand{\thefigure}{S\arabic{figure}}
\section{Determination of phase boundaries}

\begin{figure}[h]
\includegraphics[clip=true,trim= 20 190 50 180,width=0.5\columnwidth]{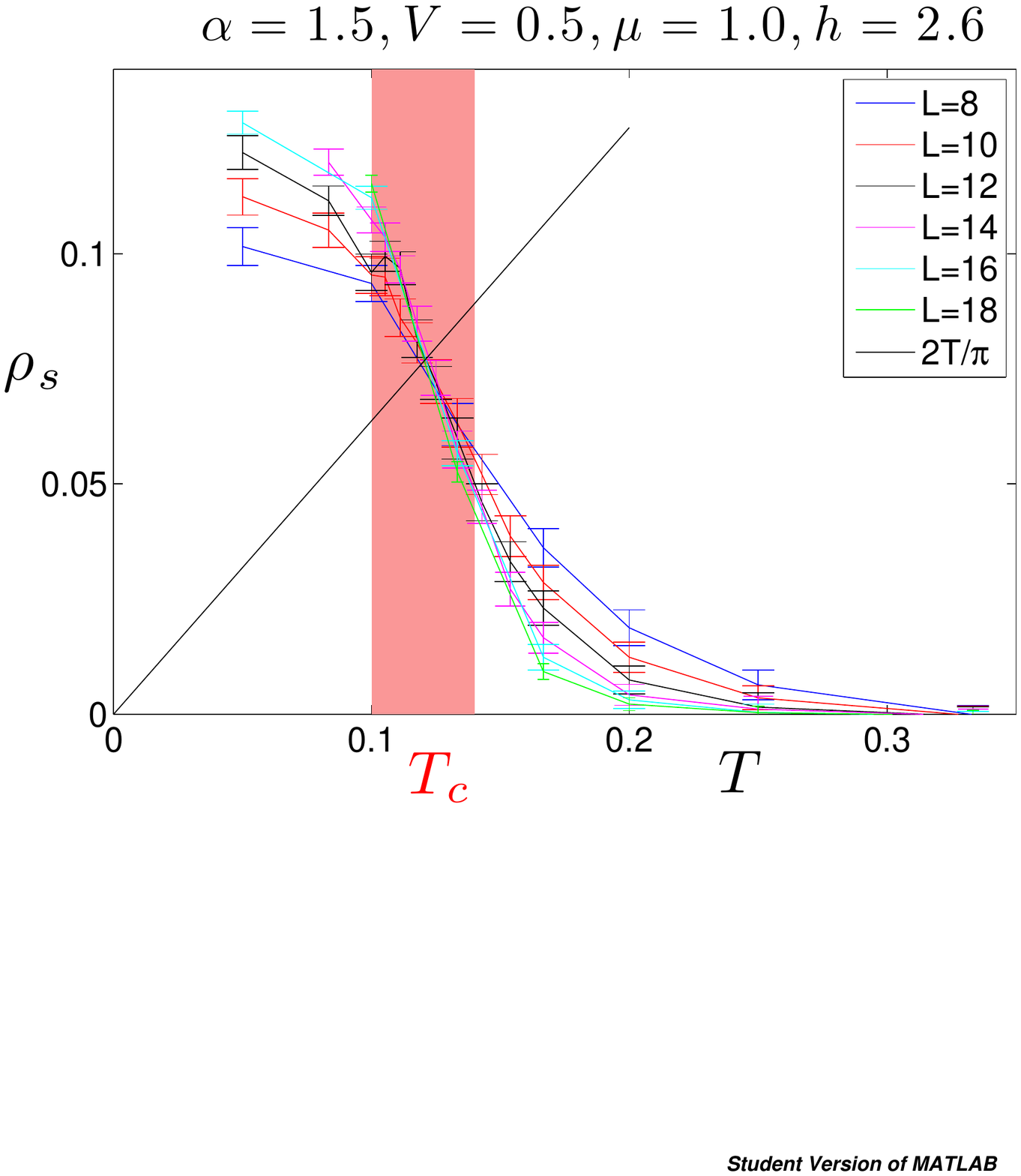}
\caption{Helicity modulus versus temperature for various system sizes. The black line represents the BKT jump of $2T/\pi$.}
\label{fig:rho_s}
\end{figure}

The nematic phase boundary has been determined using standard finite size scaling techniques appropriate to a two-dimensional classical Ising transition, as described in the appendix of Ref.  \onlinecite{Schattner2016b}. The superconducting transition temperature has been determined using the helicity modulus, following Ref. \onlinecite{Paiva2004b}:
\begin{equation}
\rho_s = \lim_{q_y\rightarrow 0} \lim_{L\rightarrow \infty} K_{xx}(q_x=0, q_y) %;L),
\label{eq:rho_s}
\end{equation}
where
\begin{equation}
K_{xx}(\mathbf{q}) \equiv \frac{1}{4} \left[\Lambda_{xx} (q_x\rightarrow 0, q_y=0)  -  \Lambda_{xx} (\mathbf{q})  \right],
\label{eq:Kxx}
\end{equation}
and $\Lambda_{xx}$ is the current-current correlator
\begin{equation}
\Lambda_{xx}(\mathbf{q}) = \sum_{i}\int_0^\beta d\tau e^{-i\mathbf{q} \cdot \mathbf{r}_i} \langle J_x(\mathbf{r}_i,\tau) J_x(0,0)\rangle.
\end{equation}
Here, the current density operator is given by $J_x(\mathbf{r}_i) = \sum_\sigma it(1 + \alpha \tau^z_{i,j}) c^\dagger_{i\sigma} c^{\vphantom{\dagger}}_{j\sigma} + \mathrm{H.c.}$,  where $\mathbf{r}_j = \mathbf{r}_i + \hat{x}$. The $q \to 0$ limits above are not strictly well defined for finite size systems, so we use the smallest nonzero momentum $q=2\pi/L$ to define a value of $\rho_s$ in finite size systems. Fig. \ref{fig:rho_s} %plots
exhibits curves of $\rho_s(T)$ for various system sizes, which cross at a temperature near where they attain the BKT jump of $2T/\pi$. Rough error bars are determined by inspection.

\begin{figure}[h]
\includegraphics[clip=true,trim= 50 260 100 280,width=0.8\columnwidth]{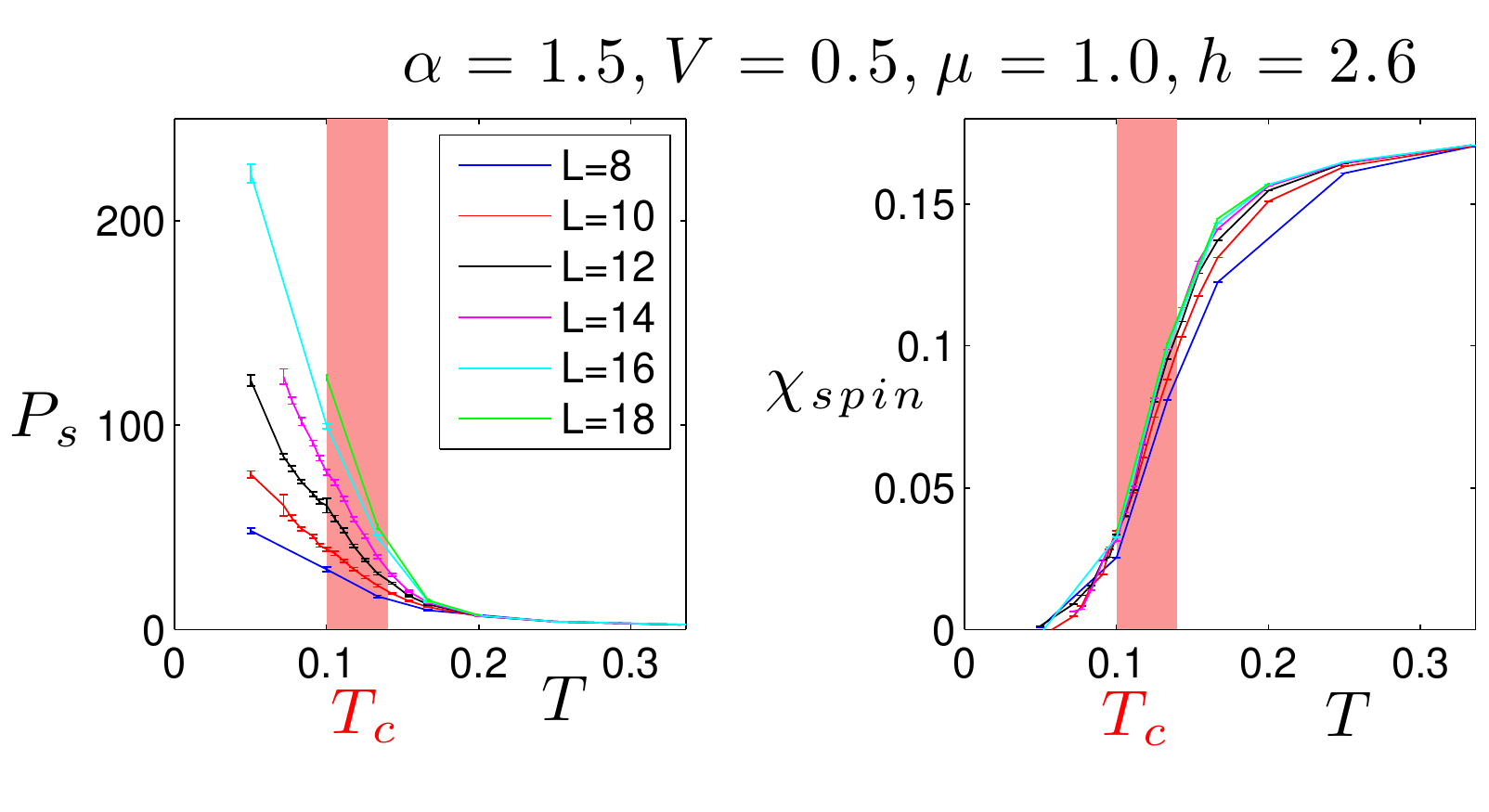}
\label{fig:other}
\caption{Left: the onset $s$-wave pairing susceptibility versus temperature. Right: the uniform spin susceptibility versus temperature. Both are shown for various system sizes, and indicate a superconducting $T_c$ consistent with that determined above.}
\end{figure}

The $T_c$ determined above can be %roughly
 corroborated by examining other thermodynamic quantities (Fig.  \ref{fig:other}). For instance %examining
  the pair susceptibility %, which diverges in the thermodynamic limit only in the superconducting phase:
\begin{equation}
P_{s} = \int_0^\beta d\tau \sum_i \langle \Delta_{s}^\dagger (\mathbf{r}_i, \tau)  \Delta^{\vphantom{\dagger}}_{s}(0,0) \rangle.
\label{eq:sc_suscept}
\end{equation}
where, $\Delta_s(\mathbf{r}_i) = c_{i\uparrow} c_{i\downarrow}$ diverges in the thermodynamic limit only in the superconducting phase. The uniform spin susceptibility also %shows the development of
shows evidence of the opening a spin gap %due to superconductivity
in the superconducting phase.

\section{estimating the DC resisitivity}

\subsection{Resistivity proxy from the long imaginary time data}

%In this part of the Supplementary Information,
Next, we describe the ``resistivity proxy'' $\rho_{2}$ introduced in the main text as an estimator for the d.c. resistivity, $\rho_{\mathrm{dc}}$. $\rho_{2}$ can be computed directly from imaginary time data (without the need for any analytical continuation). It is proportional to $\rho_{\mathrm{dc}}$ under certain assumptions, as we discuss below. We have used the resistivity proxy analysis as a complement to %the
direct analytical continuation. The fact that both approaches give a qualitatively similar temperature dependence for the resistivity, and even the quantitative estimates agree within a factor of $\sim 2$ (see Fig.~5 of the main text), provides support for the validity of our assumptions.

The imaginary-time (and Matsubara frequency) current-current correlations
are related to the real part of the real-frequency conductivity by
\begin{equation}
\Lambda(i\omega_{n})=\int\frac{d\omega}{\pi}\frac{\omega^{2}\sigma'(\omega)}{\omega^{2}+\omega_{n}^{2}}.\label{eq:omega}
\end{equation}
\begin{equation}
\tilde{\Lambda}(\tau)=\int\text{\ensuremath{\frac{d\omega}{2\pi}}}\sigma'(\omega)\frac{\omega\cosh\left[(\frac{\beta}{2}-\tau)\omega\right]}{\sinh\left(\frac{\beta\omega}{2}\right)},\label{eq:tau}
\end{equation}
{[}Note that Eq. (\ref{eq:tau}) is valid for $0\le\tau\le\beta.${]}
%We are interested in inverting these relations, finding $\sigma(\omega)$ [and, in particular, we are interested in $\sigma_{\mathrm{dc}} = \sigma(\omega=0)$].
Since the kernels in Eqs. (\ref{eq:tau},\ref{eq:omega}) are ill-conditioned, %this
inverting these equations is a highly numerically unstable problem. To make
matters worse, $\sigma'(\omega)$ turns out to have
features whose characteristic width is of order $T$ or less - which
is less than the intrinsic ``resolution'' of the kernel.  %Therefore, we have to resort to assumptions about $\sigma(\omega)$ in order to make progress.

However, under many circumstances, we would expect that $\sigma'(\omega,T)$ at low frequencies is determined by the behavior of $\tilde \Lambda(\tau)$ at the longest imaginary times, $\tau \sim \beta/2$, or equivalently by the low-frequency
moments (LFMs)~\cite{Trivedi1995b}
\begin{eqnarray}
m_{0} & \equiv & \beta\widetilde{\Lambda}(\tau=\frac{\beta}{2})=\beta\int\frac{d\omega}{2\pi}\frac{\omega\sigma'(\omega)}{\sinh\left(\frac{\beta\omega}{2}\right)},\\
m_{2} & \equiv & \beta\partial^2_{\tau}{\widetilde{\Lambda}}(\tau=\frac{\beta}{2})=\beta\int\frac{d\omega}{2\pi}\frac{\omega^{3}\sigma'(\omega)}{\sinh\left(\frac{\beta\omega}{2}\right)},
\end{eqnarray}
in terms of which the resistivity proxy defined in the text is
\begin{equation}
\rho_{2}\equiv\frac{m_{2}}{2\pi Tm_{0}^{2}}=\left.\frac{\partial^2_\tau{\widetilde\Lambda}}{2\pi\widetilde\Lambda^2}\right|_{\tau=\beta/2}
\end{equation}

To illustrate the usefulness of this definition, consider the simple case in which $\sigma'(\omega)$ is a Lorentzian,
%\begin{equation}
\begin{equation}
\sigma'(\omega,T)=\frac 1 {\rho}\left[\frac{\Gamma^2}{\omega^{2}+\Gamma^{2}}\right] \ \ \rightarrow \ \ \frac {\rho_{2}}\rho=%\frac{\pi\Gamma}{D}
\frac{2\pi^{2}-\beta\Gamma F(\beta\Gamma)}{[F(\beta\Gamma)]^{2}} %\;\textrm{[Lorentzian \ensuremath{\sigma_{D}(\omega)}]}.
\end{equation}
%Here,
where
\begin{equation}
F(x)=2x\left[\psi\left(\frac{x}{2\pi}\right)-\psi\left(\frac{x}{4\pi}\right)-\log(2)\right]-2\pi,
\end{equation}
%where
and $\psi(x)$ is the digamma function. %If, at low temperature, $\Gamma(T)\ll T$, we get that {[}using $\psi(x)=-\frac{1}{x}+\dots${]},
It is easy to see that $\rho_{2}/\rho$ does not depend very strongly on $\beta\Gamma$.  Specifically, $\rho_{2} \to \rho$ as $\beta\Gamma\to \infty$ and $\rho_{2} \to  \rho/2$ as $\beta \Gamma \to 0$ ({\it i.e.} for a narrow Lorentzian).

\begin{figure}[t]
\includegraphics[clip=true,trim= 40 190 50 180,width=0.5\columnwidth]{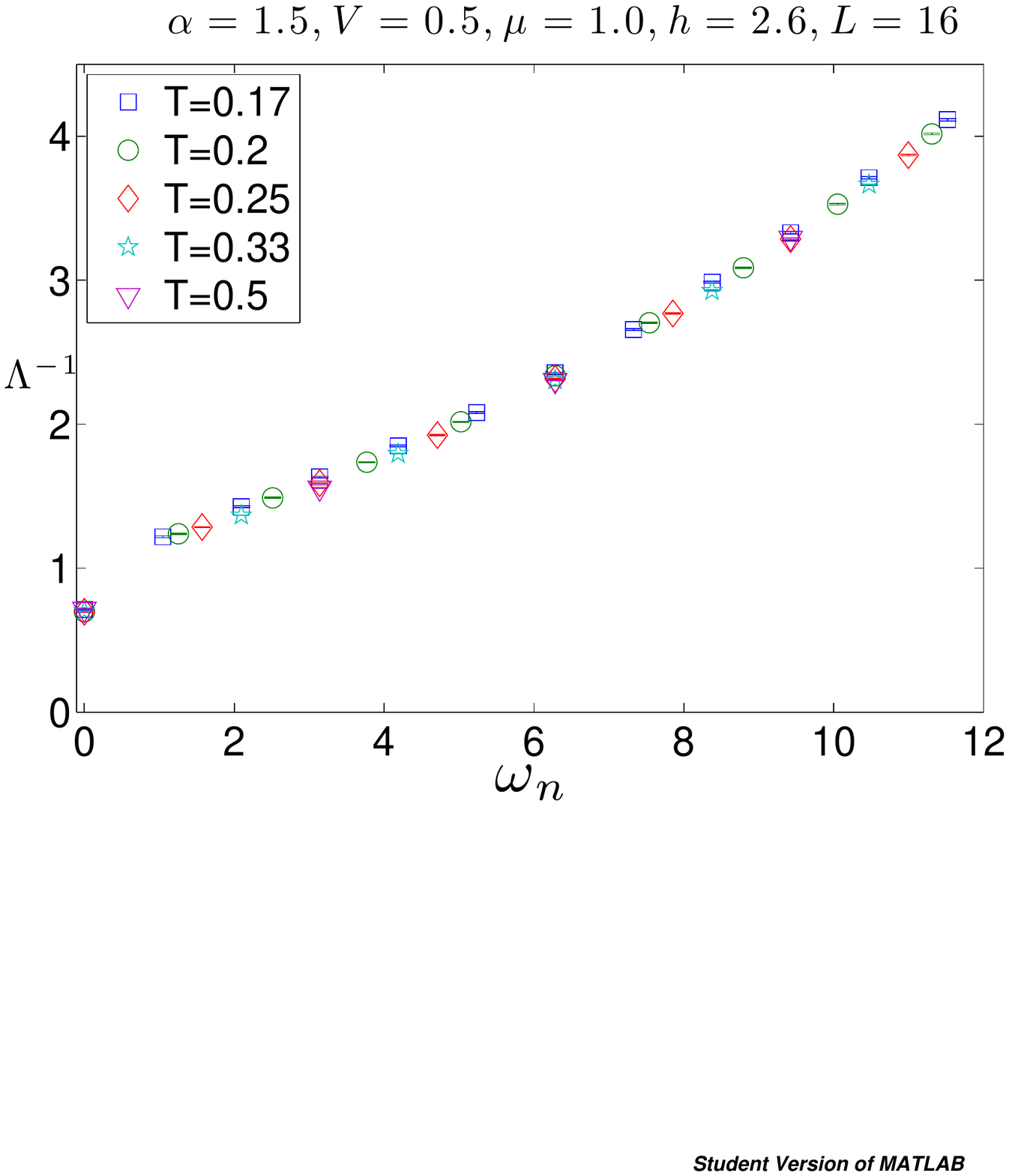}\caption{$\Lambda^{-1}(i\omega_{n})$ at $h\approx h_c$ for different temperatures. The parameters used in the runs are $\alpha=1.5$, $V=0.5$, $\mu=-1$, and $h=2.6$.
}
\label{fig:Lambda_omega}
\end{figure}

Naturally, especially at criticality, it is not reasonable to expect $\sigma'(\omega)$ to have a simple Lorentzian form.
However, $\rho_{2}$ turns out to provide a reasonable estimate of $\rho$ under much more general circumstances.
Specifically, let us assume that
\begin{equation}
\sigma'(\omega,T)=\sigma_{D}(\omega,T)+\sigma_{\mathrm{reg}}(\omega,T),\label{eq:assumption}
\end{equation}
where $\sigma_{D}(\omega,T)$ is a Drude-like piece ({\it i.e.} maximal at $\omega=0$), that satisfies
\begin{equation}
\sigma_{D}(\omega,T)\xrightarrow[T\rightarrow0]{}D\delta(\omega),
\end{equation}
and whose characteristic width at finite temperature is of the order
of $T$ or less, while $\sigma_{\mathrm{reg}}(\omega,T)$ is a regular
piece %with a non-singular $T\rightarrow0$ limit,
with a width that is always large compared to $T$ %.  $\sigma_{\mathrm{reg}}(\omega,T=0)$.
and correspondingly a magnitude at low frequencies ($\omega \lesssim T$) that is small compared to $\sigma_D$ ({\it i.e.} $\sigma'_D(0,T) \gg \sigma_\mathrm{reg}(0,T)$).
%Eq. (\ref{eq:assumption}) is what we usually associate with metallicconductivity in the clean limit.
Evidence for the validity of Eq.
(\ref{eq:assumption}) in our problem can be seen by comparing $\Lambda(i\omega_{n})$
for different temperatures. This is shown in Fig. \ref{fig:Lambda_omega}.
For all temperatures, $\Lambda(i\omega_{n})$ has an apparent ``jump''
from $\omega_{n}=0$ to $\omega_{n}>0$. $\Lambda(i\omega_{n})$ for
different temperatures are seen to approximately lie on a single,
nearly temperature-independent curve. Both features can be readily understood
from Eq. (\ref{eq:assumption}). The jump at $\omega_{n}=0$ is a
consequence of $\sigma_{D}(\omega,T)$, which %is essentially almost
so long as its width is less than $2\pi T$ behaves effectively as if it were
a delta function at $\omega_{n}=0$. % (since its width is smaller than$T$).
The fact that finite $\omega_n$ data from different temperatures lie on a single
curve %is a manifestation of the convergence of
 suggests that $\sigma_{\mathrm{reg}}(\omega,T)$ is not strongly $T$ dependent over the relevant range of $T$.
%at low temperature.

%\subsection{Low-frequency moments and resistivity proxy}

%In principle, $\sigma_{\mathrm{reg}}(\omega)$
%can be estimated using standard analytic continuation techniques,
%such as the maximum entropy method. These techniques can hardly give
%any information on $\sigma_{D}(\omega)$, since it is so narrow.
%To extract information about $\sigma_{D}(\omega)$, we compute the low-frequencymoments (LFMs)~\cite{Trivedi1995b}
%\begin{eqnarray}
%m_{0} & \equiv & \beta\tilde{\Lambda}(\tau=\frac{\beta}{2})=\beta\int\frac{d\omega}{2\pi}\frac{\omega\sigma(\omega)}{\sinh\left(\frac{\beta\omega}{2}\right)},\\
%m_{2} & \equiv & \beta\ddot{\tilde{\Lambda}}(\tau=\frac{\beta}{2})=\beta\int\frac{d\omega}{2\pi}\frac{\omega^{3}\sigma(\omega)}{\sinh\left(\frac{\beta\omega}{2}\right)}.
%\end{eqnarray}

\begin{figure}[t]
\includegraphics[clip=true,trim= 20 230 30 230,width=0.9\columnwidth]{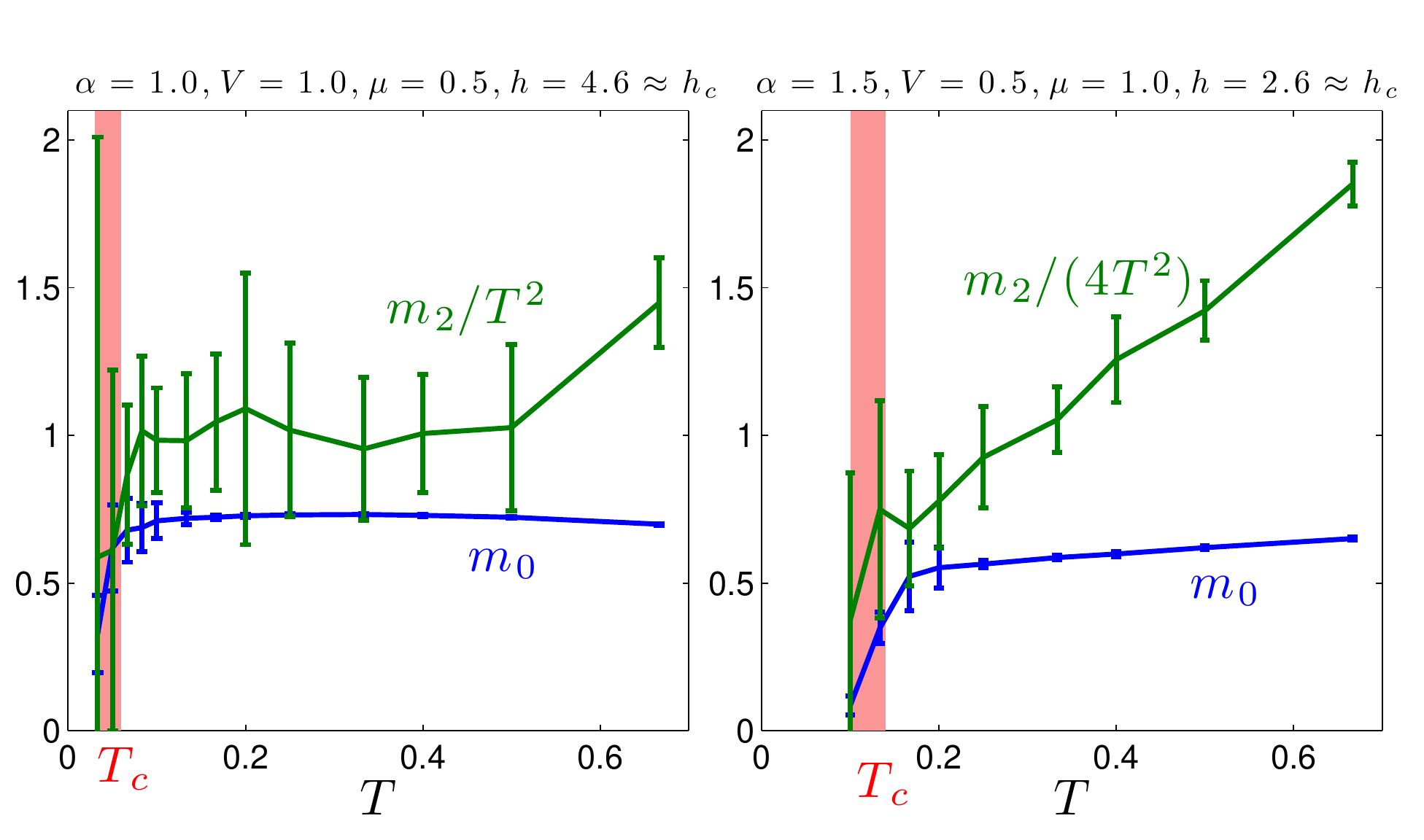}
\caption{$m_{0}$ and $m_{2}/T^{2}$ as a function of $T$ at criticality.}
\label{fig:beta_over_2}
\end{figure}

%We did not dare to consider higher moment, as the error bars in these are probably too high.
We are now faced with the task of disentangling $\sigma_{\mathrm{reg}}(\omega)$
and $\sigma_{D}(\omega)$.
The LFMs clearly combine information about
both $\sigma_{\mathrm{reg}}$ and $\sigma_{D}$. Crucially, however,
the two contributions \emph{may scale differently with temperature
as $T\rightarrow0$}. Consider $\sigma_{\mathrm{reg}}$: its contribution
to the LFMs, to lowest order in $T$, is expected to behave as
\begin{eqnarray}
m_{0,\mathrm{reg}} & = & \beta\int\frac{d\omega}{2\pi}\frac{\omega\left[\sigma_{\mathrm{reg}}(0)+\frac{1}{2}\partial_\omega^2 \sigma_{\mathrm{reg}}(0)\omega^{2}+\dots\right]}{\sinh\left(\frac{\beta\omega}{2}\right)}\sim T+O(T^{3}),\nonumber \\
m_{2,\mathrm{reg}} & = & \beta\int\frac{d\omega}{2\pi}\frac{\omega^{3}\left[\sigma_{\mathrm{reg}}(0)+\frac{1}{2} \partial_\omega^2 \sigma_{\mathrm{reg}}(0)\omega^{2}+\dots\right]}{\sinh\left(\frac{\beta\omega}{2}\right)}\sim T^{3}+O(T^{5}).
\end{eqnarray}
In contrast, %since
assuming $\sigma_{D}(\omega)$ has a width of the order
of $T$ or less, its corresponding LFMs %may have a different scaling.
will scale differently.
If the total weight of $\sigma_{D}(\omega)$ has a non-zero limit
as $T\rightarrow0$, and if the characteristic width of $\sigma_{D}(\omega)$
is much less than $T$, we expect
\begin{equation}
m_{0,D}=\beta\int\frac{d\omega}{2\pi}\frac{\omega\sigma_{D}(\omega)}{\sinh\left(\frac{\beta\omega}{2}\right)}\approx\int\frac{d\omega}{\pi}\sigma_{D}(\omega),
\end{equation}
and hence, this is non-zero in the $T\rightarrow0$ limit. This is
true also if the characteristic width of $\sigma_{D}(\omega)$ is
of the order of $T$. For example, if $\sigma_{D}$ has a scaling
form: $\sigma_{D}(\omega,T)=\frac{1}{T}f\left(\frac{\omega}{T}\right)$,
then we get that $m_{0,D}\rightarrow\int_{-\infty}^{\infty}dx\frac{xf(x)}{2\sinh\left(\frac{x}{2}\right)}=\mathrm{const.}$
in the limit $T\rightarrow0$. In our data, we find that over a range
of temperatures, $m_{0}(T)$ is weakly temperature dependent,
suggesting that it is dominated by $m_{0,D}$ (see Fig.~\ref{fig:beta_over_2}).

The interpretation of $m_{2}$ is more subtle, as it depends
on the precise form of $\sigma_{D}(\omega)$ at frequencies of the
order of $T$. % or less.
 It is instructive to consider a simple model
for $\sigma_{D}(\omega,T)$:

\begin{equation}
\sigma_{D}(\omega,T)=\begin{cases}
\sigma_{0}(T), & \omega<\omega_{0}(T),\\
\sigma_{0}(T)\left(\frac{\omega}{\omega_{0}}\right)^{-\alpha}, & \omega_{0}(T)<\omega.
\end{cases}\label{eq:model_sigma}
\end{equation}
with $\omega_{0}(T)<AT$, where $A$ is a constant. Having a finite
optical weight requires $\alpha>1$. Then, to be consistent with the
observation that $m_{0,D}\approx\mathrm{const.}$, we require
that $\sigma_{0}(T)\omega_{0}(T)=D=\mathrm{const.}$ We then get that
\begin{eqnarray}
m_{2,D} & \approx & \sigma_{0}(T)\int_{0}^{\omega_{0}(T)}d\omega\omega^{2}+\sigma_{0}\int_{\omega_{0}(T)}^{T}d\omega\omega^{2}\left(\frac{\omega}{\omega_{0}}\right)^{-\alpha}\nonumber \\
 & = & \frac{1}{3}D\omega_{0}^{2}+\frac{1}{3-\alpha}D\omega_{0}^{2}\left[\left(\frac{T}{\omega_{0}}\right)^{3-\alpha}-1\right].
\end{eqnarray}
We see that, if $\omega_{0}/T\rightarrow0$ as $T\rightarrow0$, we
get that $m_{2,D}/T^{2}\rightarrow0$ as $T\rightarrow0$. For
example, if $\omega_{0}\sim T^{1+\varepsilon}$ with $\varepsilon>0$,
then $m_{2,D}\sim T^{\mathrm{min}[2+(\alpha-1)\varepsilon,\,2(1+\varepsilon)]}$.
On the other hand, if $\omega_{0}\sim T$, we get that $m_{2,D}\sim T^{2}$.

As %a concrete example,
another example, one can analyze the low-frequency conductivity
of a clean Fermi liquid with umklapp scattering. The conductivity
is given by
\begin{equation}
\sigma_{\mathrm{FL}}(\omega)=\frac{D}{\pi}\frac{\Gamma_{\mathrm{tr}}(T,\omega)}{[\Gamma_{\mathrm{tr}}(T,\omega)]^{2}+\omega^{2}},
\end{equation}
where the Fermi liquid transport scattering rate is $\Gamma_{\mathrm{tr}}(T,\omega)=(\omega^{2}+4\pi^{2}T^{2})/W$
($W$ is of the order of the Fermi energy). The integral for $m_0$ can be calculated in the limit $T\rightarrow 0$, and gives $m_0 = D$. The integral for $m_2$ is dominated by frequencies $\omega\sim T$, %and we find that
such that at low temperatures, $%m_0
m_2\sim T^3$.

%\textcolor{red}{.... Work out FL example ....}

Turning to our DQMC data, $m_{0}$ and $m_{2}$ as a function of temperature at criticality are shown in Fig. \ref{fig:beta_over_2} for two sets of parameters.
We find that both $m_{0}$ and $m_{2}/T^{2}$ extrapolate to a non-zero value in the limit $T \rightarrow 0$.
%EBa constant upon lowering $T$.
%
This implies that these moments are indeed dominated by the contribution of $\sigma_D$.  For the larger $T_c$ parameter set (right panel), both
quantities show a sudden drop at temperatures below $T=0.2t$; this is likely to be an effect of superconducting
fluctuations upon approaching the superconducting critical temperature,
$T_{c}$$\approx0.11t$. Over a range of temperatures above $T=0.05t$ (left panel) and $T=0.2t$ (right panel), the observed behavior is consistent with $\omega_{0}\propto T$ in
Eq. (\ref{eq:model_sigma}) and with $\sigma_{0}\propto1/T$.

%Assuming that the main contribution for $m_{0}$, $m_{2}$ comes from $\sigma_{D}(\omega)$, we define the ``resistivity proxy,''

%This quantity has units of resistivity, and carries information about $\sigma_{D}$.

It is worth noting that there are other possible ways to define a resistivity proxy in terms of $m_0$ and $m_2$, depending on the assumed form of $\sigma$ at criticality.
%At a metallic quantum critical point, there is no accepted theory for the optical conductivity. In particular, there is no guaranteethat it has a Lorentzian form.
 For instance, if $\sigma_{D}(\omega)=\sigma_{0}g\left(\frac{\omega}{\Gamma}\right)$,
where $g(x)$ is a dimensionless function with a well-defined second
moment (unlike a Lorentzian) and $\Gamma\le AT$, we get that $m_{0}\propto\sigma_{0}\Gamma$
and $m_{2}\propto\sigma_{0}\Gamma^{3}$. In this case, a more
appropriate resistivity proxy is

\begin{equation}
\tilde\rho_{2}=\frac{\sqrt{m_{2}/m_{0}}}{m_{0}}\propto\frac{1}{\sigma_{0}}.
\end{equation}
In the particular case where $\Gamma\propto T$, $\sigma_{0}\propto1/T$,
both $\rho_{2}$ and $\tilde\rho_{2}$ are proportional to $T$.
%Interestingly, over a range of temperature, our data is consistent with such behavior.
Reassuringly, computing  $\tilde\rho_{2}$ from our DQMC data at criticality  produces qualitatively similar results as $\rho_{2}$.

\subsection{Fitting function for $\Lambda(\omega_n)$}

In order to fit the imaginary-time data for the current-current correlator, $\Lambda(\omega_n)$, we use the following model for the optical conductivity [Eq.~(4) of the main text]:
\begin{eqnarray}
 \Lambda_{\text{fit}}(\omega_n)=\sum_{j=1}^N %\Lambda_j(\omega_n), \nonumber \\
%&& \Lambda_j(\omega_n)=
\frac{A_j}{\omega_n^2+\gamma_j|\omega_n|+\Omega_j^2},
\label{eq:lambda_fit_mats_supp}
\end{eqnarray}
with free parameters $A_j$, $\gamma_j$, $\Omega_j$, and we have found that $N=2$ is sufficient to fit our data. When comparing this form to the QMC data, we need to recall that the QMC simulations are performed with a finite imaginary time step, $\Delta \tau$.
The imaginary time correlation function $\tilde{\Lambda}(\tau)$ is sampled at discrete values $\tau = n \Delta \tau$ (where $n$ is an integer), and $\Lambda(\omega_n)$ is its discrete Fourier transform.
In particular, $\Lambda(\omega_n)$ is %is
periodic in $\omega_n$ with a period of $2\pi / \Delta \tau$. The discrete imaginary time version of (\ref{eq:lambda_fit_mats_supp}), which is appropriate for comparison with our Matsubara frequency data, is
\begin{align}
\Lambda_{\mathrm{fit}, \Delta \tau}(\omega_{n}) & =\sum_{j=1}^{N}\sum_{q=-\infty}^{\infty}\frac{A_j}{\left(\omega_{n}-\frac{2\pi q}{\Delta\tau}\right)^{2}+ \gamma_j |\omega_{n}-\frac{2\pi q}{\Delta\tau}| + \Omega_j^2}.
\end{align}
The sum over $q$ can be performed explicitly, using
\begin{align}
\sum_{q=-\infty}^{\infty}\frac{1}{(x+q)^{2}+A|x+q|+B^{2}} & =\frac{\psi\left(\frac{A}{2}+\frac{\sqrt{A^{2}-4B^{2}}}{2}+x\right)-\psi\left(\frac{A}{2}-\frac{\sqrt{A^{2}-4B^{2}}}{2}+x\right)}{\sqrt{A^{2}-4B^{2}}}\nonumber \\
 & +\frac{\psi\left(1+\frac{A}{2}+\frac{\sqrt{A^{2}-4B^{2}}}{2}-x\right)-\psi\left(1+\frac{A}{2}-\frac{\sqrt{A^{2}-4B^{2}}}{2}-x\right)}{\sqrt{A^{2}-4B^{2}}},
\end{align}
where $\psi(x)$ is a polygamma function, $A = \frac{\gamma_j \Delta \tau}{2\pi} $, $B = \frac{\Omega_j \Delta \tau}{2\pi}$, and $x = \frac{\omega_n \Delta \tau}{2\pi}$ ($0\le x<1$). This is the form we used in our fits to the QMC data.

\subsection{Sources of error for transport measurements}
\begin{figure}[h]
\includegraphics[clip=true,trim= 50 260 100 250,width=0.7\columnwidth]{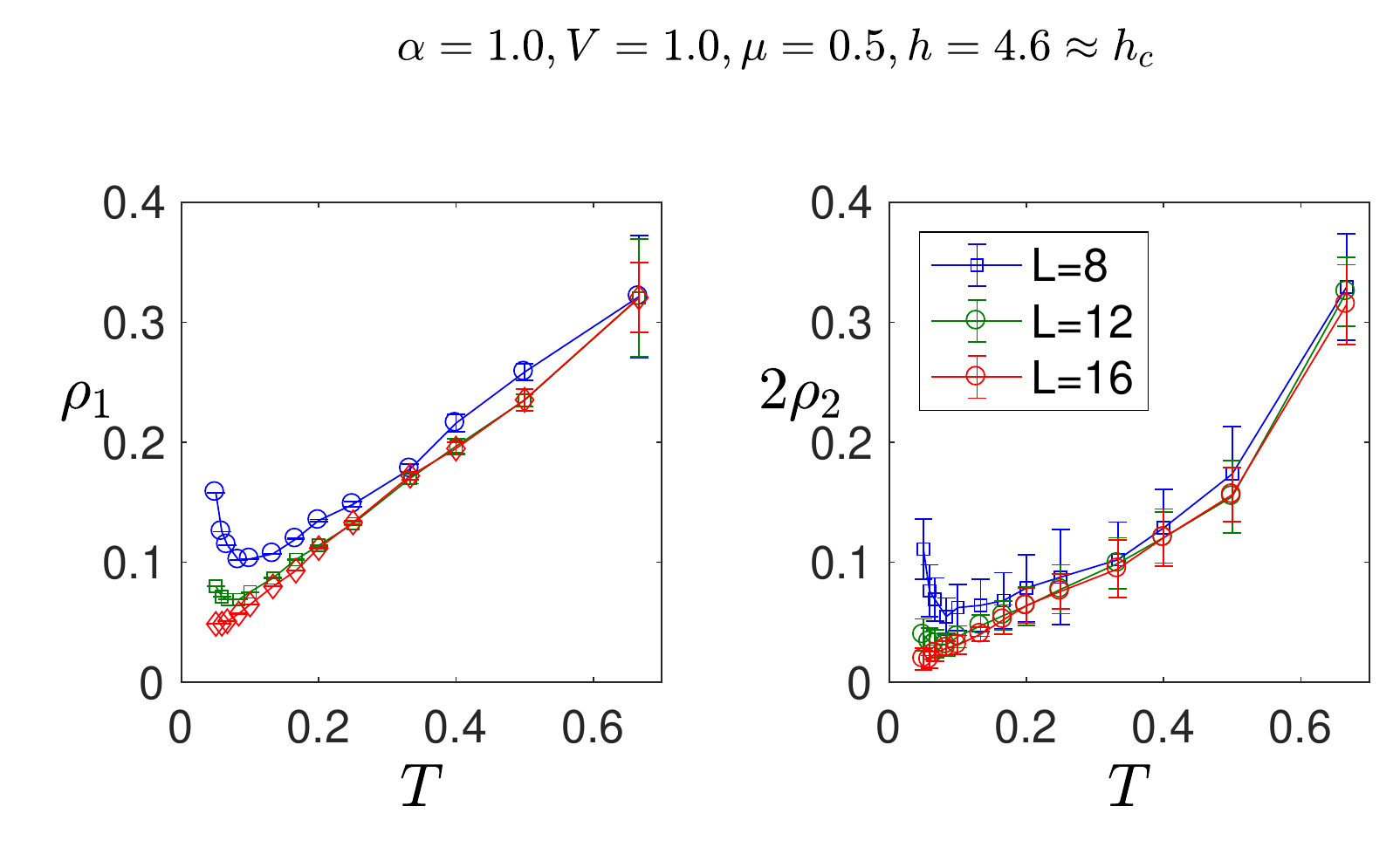}\caption{The temperature dependence of $\rho_{1}$ and $\rho_{2}$ for various system sizes, including error estimates.}
\label{fig:rho}
\end{figure}

Our numerical experiments entail several sources of error, particularly with respect to measurement of the DC resistivity via $\rho_{1}$ and $\rho_{2}$.  In this appendix we first discuss error estimates for the finite size data shown in Fig. \ref{fig:rho}, and then estimate the error entailed by taking the largest available system to represent the thermodynamic limit.

$\rho_{2}$ is extracted directly from the value and second $\tau$ derivative of $\widetilde\Lambda(\tau)$ at $\tau=\beta/2$, which are in turn determined by a linear fit of $\widetilde\Lambda(\tau)$ vs $(\tau-\beta/2)^2$ over an appropriate window. The statistical errors on $\widetilde \Lambda(\tau)$ give rise to straightforward confidence intervals on $\widetilde\Lambda(\beta/2)$ and $\partial_\tau^2{\widetilde\Lambda}(\beta/2)$, which are propagated to yield the error estimates for $\rho_{2}$ reflected in Fig. \ref{fig:rho}.

\begin{figure}[h]
\includegraphics[clip=true,trim= 20 250 50 250,width=0.7\columnwidth]{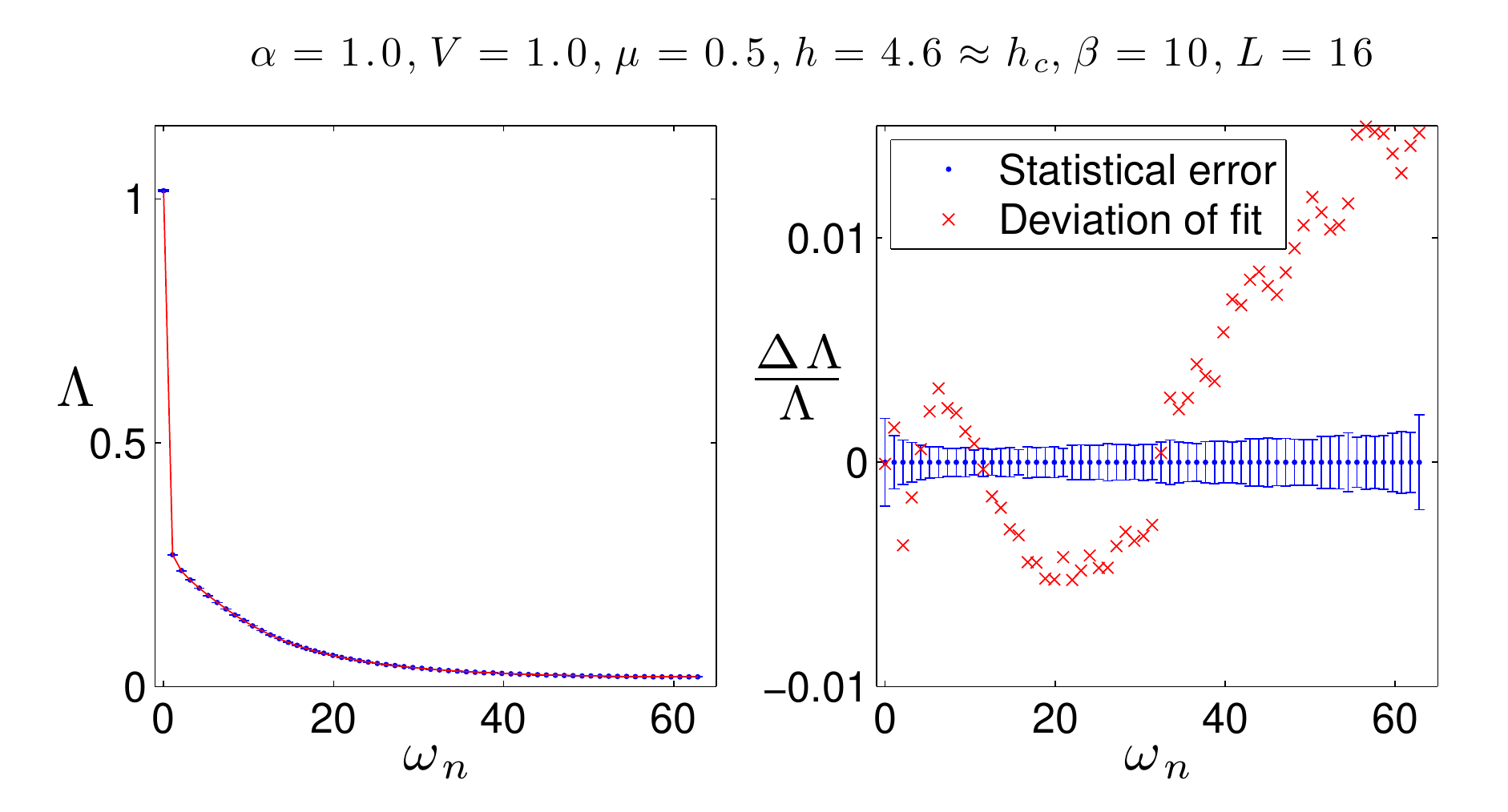}\caption{Left: data for $\Lambda(\omega_n)$ and the two-component fit described in the text. Right: the deviation between fit and data, expressed as a fraction of $\Lambda$, with statistical errors for comparison.}
\label{fig:fit_error}
\end{figure}

\begin{figure}[h]
\includegraphics[clip=true,trim= 20 240 50 240,width=0.8\columnwidth]{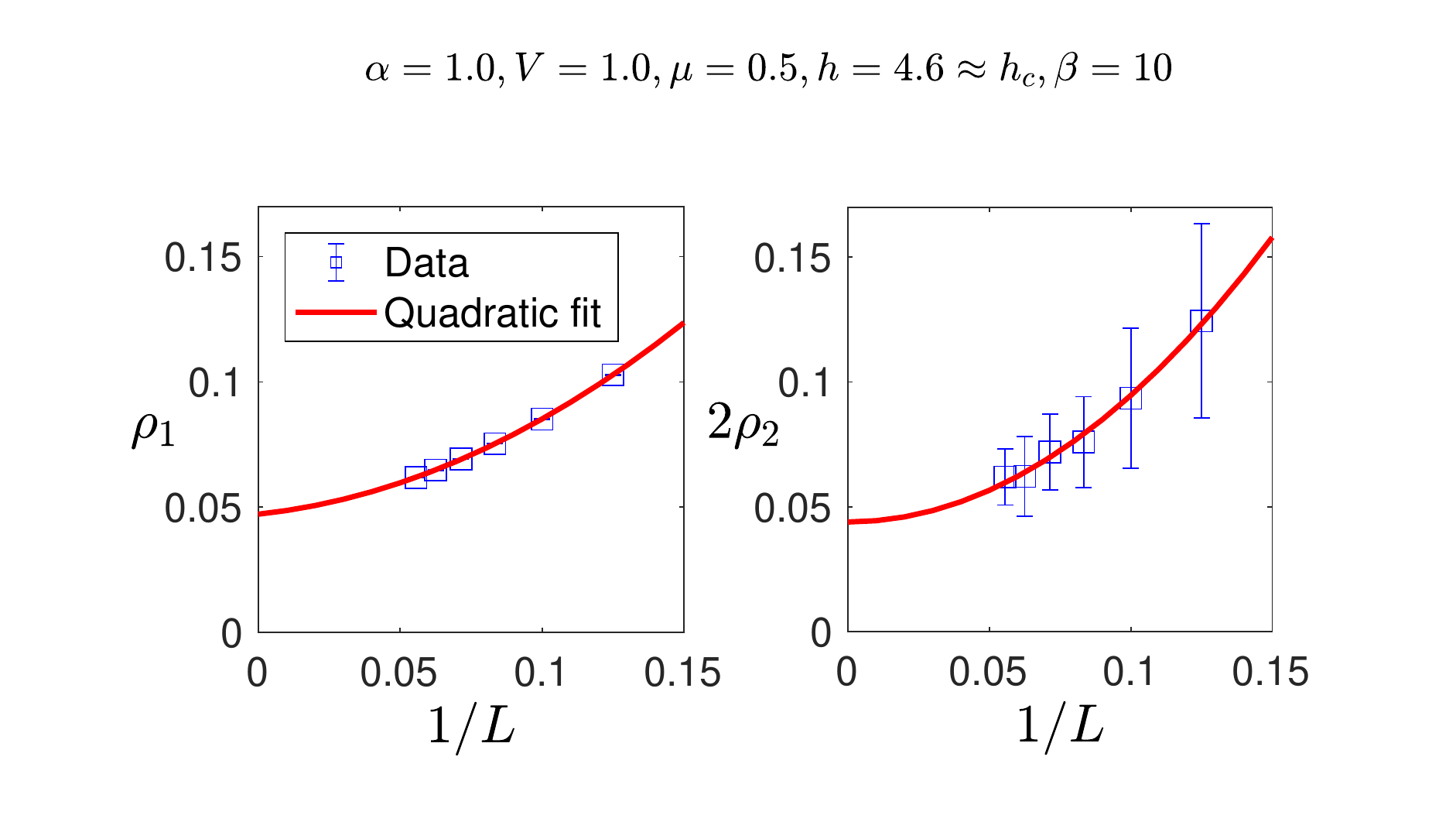}\caption{Estimation of finite size error via a quadratic fit in $1/L$.}
\label{fig:finite}
\end{figure}

\begin{figure}[h]
\includegraphics[clip=true,trim= 20 200 50 200,width=0.6\columnwidth]{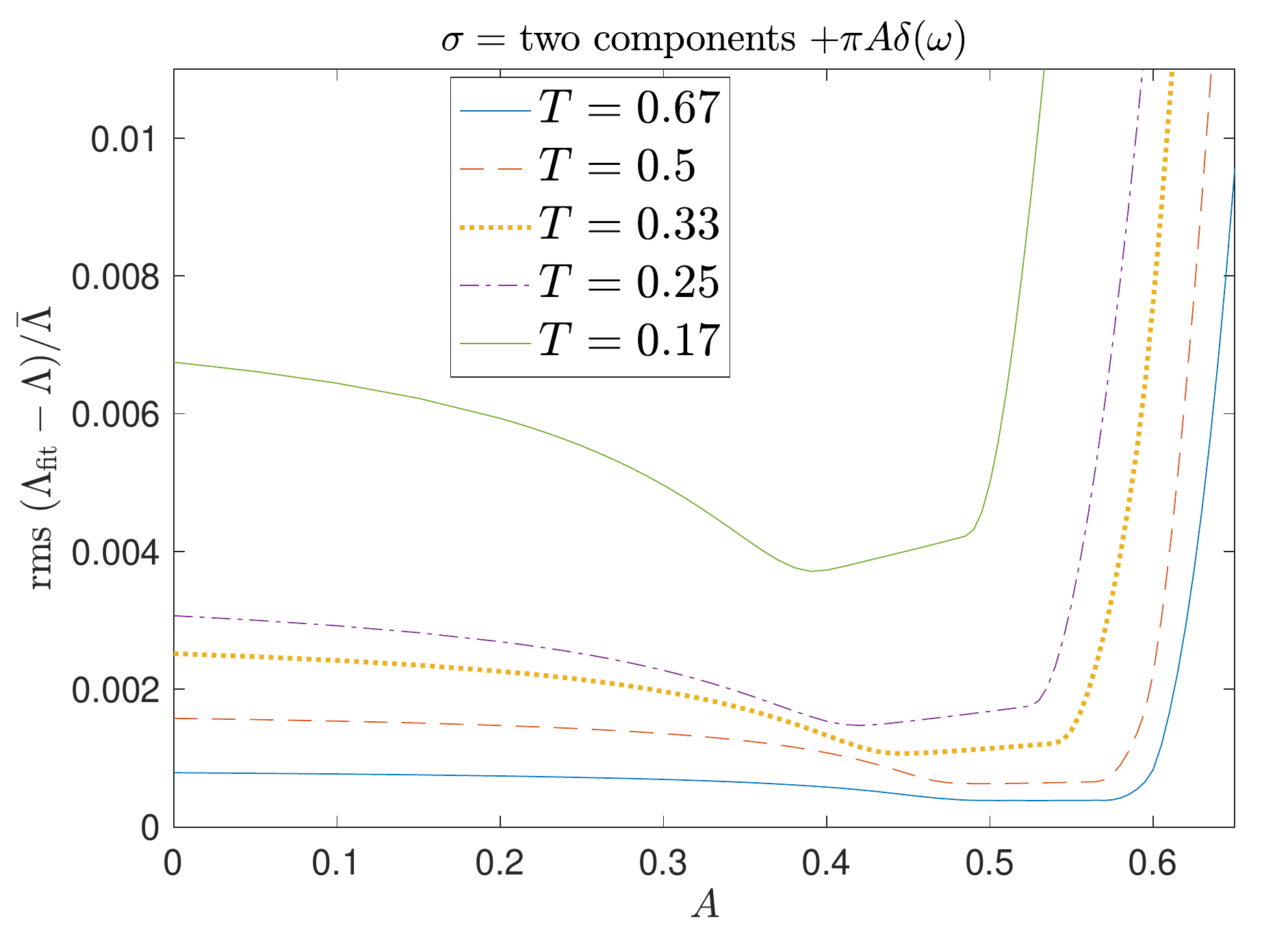}\caption{Softness of the two component conductivity fit to the addition of a sharp Drude peak. Plotted is the root mean squared deviation of the fit from the numerical data, normalized by $\bar \Lambda$, the mean value of $\Lambda$. The parameters used in the QMC run were {$\alpha=1.5$, $V=0.5$, $\mu=1$, $h=2.6\approx h_c$, and $L=20$}. The quality of the fit is modestly improved by the addition of a delta function component of fixed weight, with the remaining two components of form \eqref{eq:lambda_fit_mats}. As expected, we cannot rule out additional structure of the conductivity at frequencies far less than the temperature.}
\label{fig:delta_peak}
\end{figure}

Fig. \ref{fig:fit_error} shows that statistical error is not a meaningful source of error for $\rho_{1}$. While the two component fit described in the text is consistently within a few percent of $\Lambda(\omega_n)$, the deviation substantially exceeds the statistical error bars on $\Lambda(\omega_n)$ in a systematic, frequency dependent way. It is beyond our simple approach to estimate the magnitude of the error in $\rho_{1}$ that this systematic deviation entails. An additional source of error is in numerical minimization--different starting guesses for the least squares algorithm lead to variation of the inferred $\rho_{1}$ due to a broad minimum in the objective function, particularly at high temperature. This variation (for several choices of starting guess) yields the error bars pictured in Fig.  \ref{fig:rho}.

For both $\rho_{1}$ and $\rho_{2}$, there are systematic finite-size errors that become increasingly important at low temperature. In the text, we quote the value for the largest system size simulated (between  $16\times 16$ and $20\times 20$, depending on the temperature). To estimate the magnitude of the finite-size error, we perform a quadratic fit to the data as a function of $1/L$, as shown in Fig. \ref{fig:finite}. The difference between the $1/L\to0$ extrapolation and the value for the largest system size is our estimate for the finite size error. The finite size error and the fitting/minimization error are added in quadrature to produce the error estimates in the paper. An identical procedure of error estimation is employed for the width and weight of the Drude-like piece shown in the paper.

\subsection{Insensitivity of the fit to an additional narrow peak in $\sigma'(\omega)$}

In the main text, we have used a simple two-component form for the conductivity $\sigma'(\omega)$ to fit the Matsubara frequency data. This is the ``minimal'' form that is required, in the sense that the data cannot be well described with a single component form. However, as we now show, we cannot exclude the presence of \emph{additional} components. In particular, the data can be described equally well with an additional narrow peak in $\sigma'(\omega)$ centered at $\omega=0$, whose width is much smaller than $T$.

In order to demonstrate this, we have repeated the fitting analysis describes in the paper with an additional delta-function contribution to $\sigma'(\omega)$ with a variable weight $A$. This corresponds to the following fitting function:
\begin{equation}
\tilde{\Lambda}_{\mathrm{fit}} = \Lambda_{\mathrm{fit}} + \pi A \delta_{\omega_n, 0},
\end{equation}
where $\Lambda_{\mathrm{fit}}$ is the two component form used in the main text [Eq.~(\ref{eq:lambda_fit_mats}) above]. In Fig.~\ref{fig:delta_peak} we present the root mean square deviation of the best fit as a function of $A$. As seen in the figure, the fit quality improves slightly upon increasing $A$, until it reaches a certain critical value where the deviation turns up sharply. We conclude that our analysis of the conductivity cannot rule out the presence of additional ``fine structure'' of $\sigma'(\omega)$ at frequencies $\omega \lesssim T$. In particular, if such fine structure is present, the resistivity proxies can be dramatically different from the true DC resistivity. (For example, if $A\ne 0$, the DC resistivity is zero.)

Therefore, the resistivity proxies which we obtained by analyzing the imaginary time (or Matsubara frequency) correlator {cannot} be related to the true DC resistivity without further assumptions. The proxies can tell us about the DC resistivity only if $\sigma'(\omega)$ has a sufficiently simple structure at low frequency, such as Eq.~(\ref{eq:assumption}) above. The appeal of this form is in its simplicity; however, there are well-defined scenarios where it might fail, e.g. due to the emergence of an approximate conserved momentum in the presence of a sharp Fermi surface. Whether this is the case in our problem requires going beyond the present analysis.

\subsection{Comparison with maximal-entropy analytic continuation}
\begin{figure}[h]
    \includegraphics[width=0.9\columnwidth]{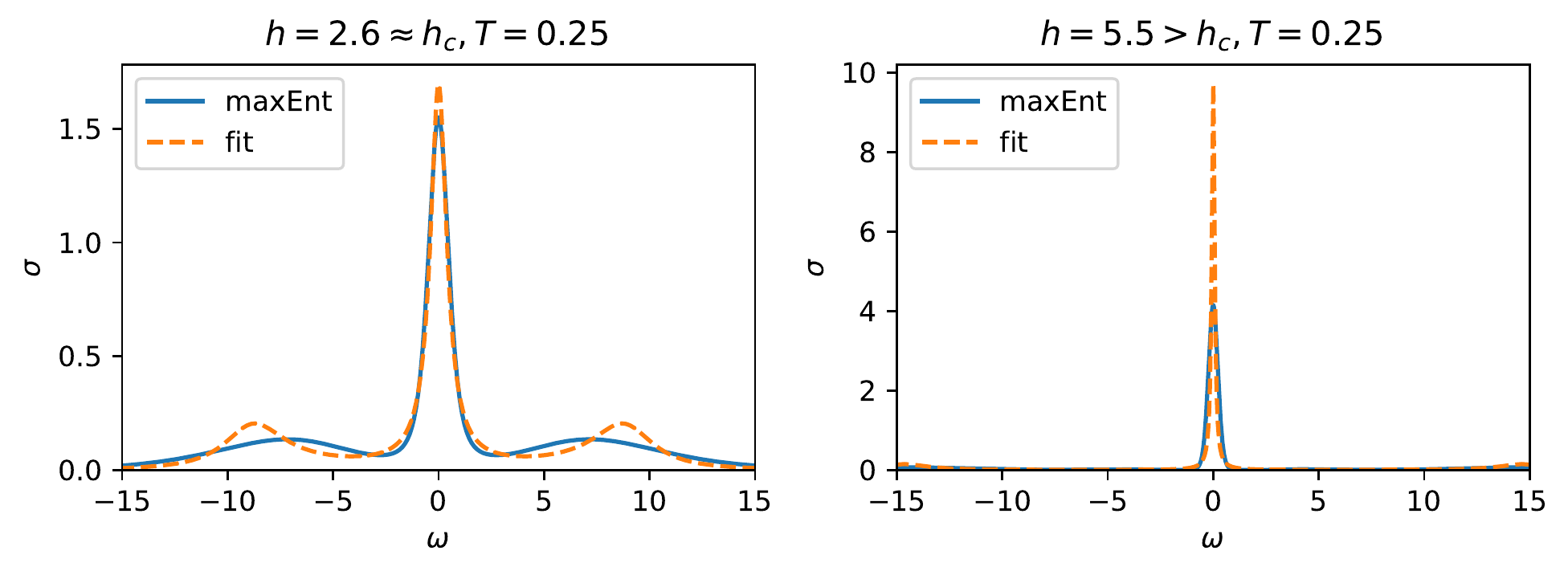} \\
    \includegraphics[width=0.9\columnwidth]{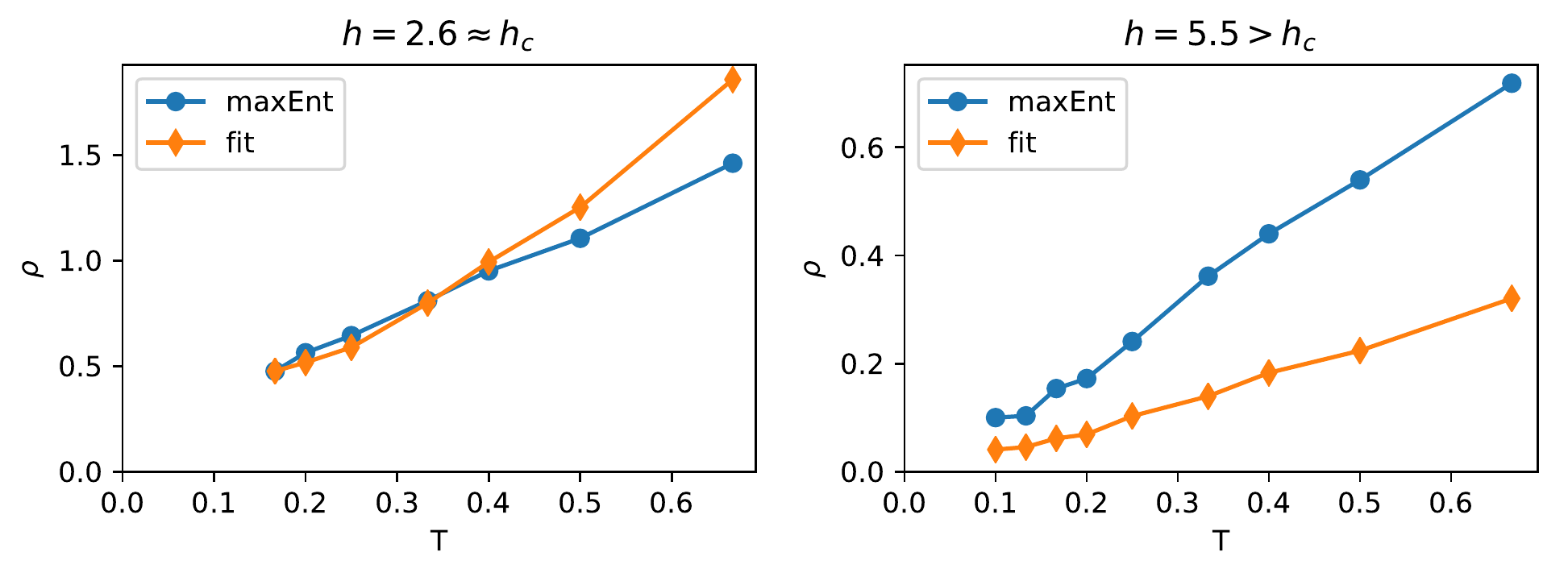}
    \caption{Top panels: The optical conductivity $\sigma(\omega)$ as obtained by maxEnt, compared with the fitting function shown in the main text.
    Bottom panels: The temperature dependence of the DC resistivity obtained by both methods. Shown here for $\alpha=1.5, V=0.5, \mu = 1$}
    \label{fig:maxent_conductivity}
\end{figure}
As an additional test on our results, we obtain the conductivity by applying the maximal entropy method (maxEnt), using Bryan's algorithm~\cite{Jarrell1996,Bryan1990}.
The maxEnt method, while not as constrained as the fitting approach we have used in the main text, is biased towards producing broad, featureless spectral functions.
The resulting conductivity $\sigma(\omega)$, shown in Fig.~\ref{fig:maxent_conductivity} has two-peak structure much like the fitting function used in the main text.
Close to criticality, the Drude-like peak at $\omega=0$ has width comparable to $T$, and the resistivity roughly matches the results of the fit, $\rho_{\mathrm{maxEnt}}\approx \rho_2$.
There is a greater discrepancy between the two component fit and the maximum entropy result far from $h_c$. This is consistent with the development of a parametrically sharp Drude peak (i.e. one with width much smaller than $T$), a feature likely to be ill-captured by any unbiased form of analytic continuation.

\section{Fermion mass enhancement}
\label{sec:mass}
We present evidence in the main text that the low temperature metallic states for $h$ away from $h_c$ are Fermi liquid-like in character, and should therefore be characterized by an effective quasiparticle dispersion. In this section we describe a method to approximately measure this dispersion near the Fermi surface, and show that it is subject to substantial flattening as $h\to h_c$. This reduction in the effective Fermi velocity, which is observed everywhere on the Fermi surface except the cold spots, is typically described as an enhanced effective mass.

We define a low-frequency moment of the spectral function,  $\Omega_1\left(\vec k\right)$ from the Fermion green function $\widetilde G\left(\vec k,\tau\right)$ according to
\begin{equation}
\Omega_1\left(\vec k\right) = -\partial_{\tau} \log\left[\widetilde G\left(\vec k,\tau\right)\right]\Bigg |_{\tau=\beta/2}=\frac{\int d\omega \frac{\omega A\left(\vec k,\omega\right)}{\cosh(\beta\omega/2)}}{\int d\omega \frac{A\left(\vec k,\omega\right)}{\cosh(\beta\omega/2)}},
\end{equation}
where the final equality is an exact identity. For a free fermion system, $\Omega_1\left(\vec k\right) $ precisely equals the dispersion. In a Fermi liquid at temperature $T$, there is a renormalized quasiparticle dispersion $\epsilon\left(\vec k\right)$, and as $\vec k$ approaches the Fermi surfaces, the spectral function is dominated by a peak centered at $\epsilon\left(\vec k\right)$, with width much less than $T$. Accordingly, $\Omega_1\left(\vec k\right)\to \epsilon\left(\vec k\right)$ as $\epsilon\left(\vec k\right)/T\to 0$. Therefore, in a Fermi liquid, $\Omega_1\left(\vec k\right) $ is a valid proxy for the dispersion within a range $\sim T$ of the Fermi level. The assumptions above clearly break down close to criticality (at least away from the cold regions), where there are no well-defined quasiparticles.

\begin{figure}[h]
\includegraphics[width=0.9\columnwidth]{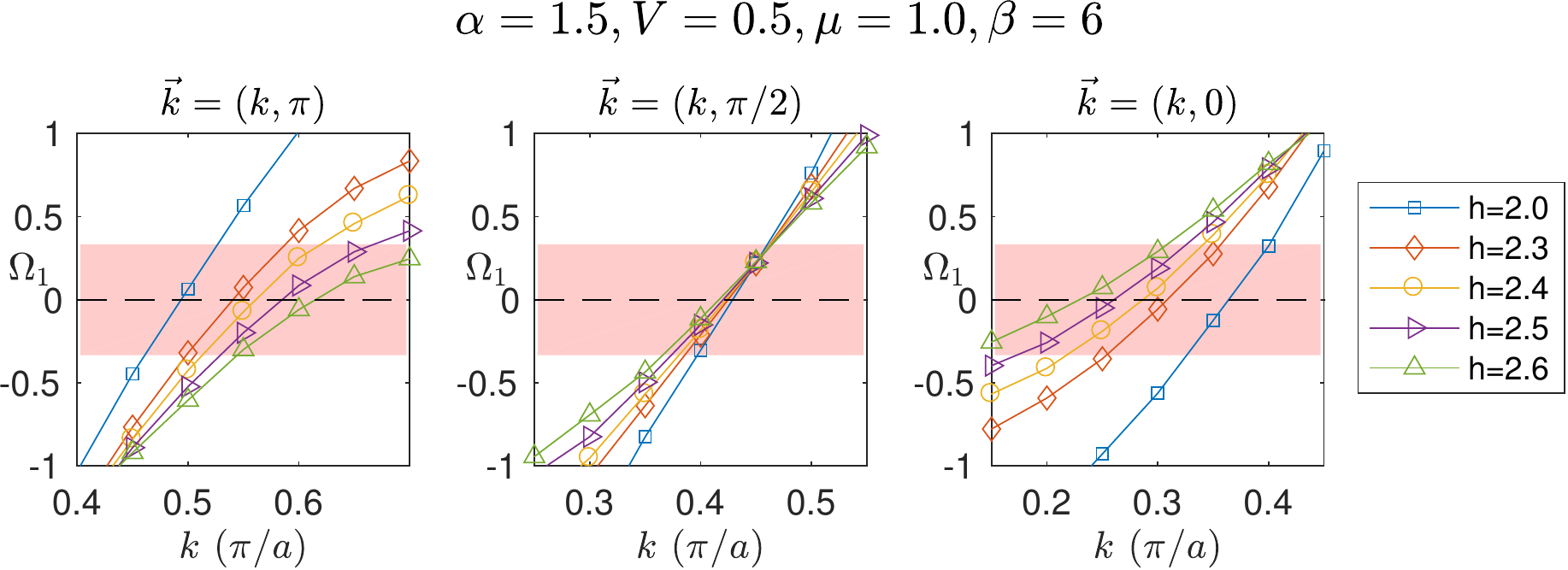}
\includegraphics[width=0.9\columnwidth]{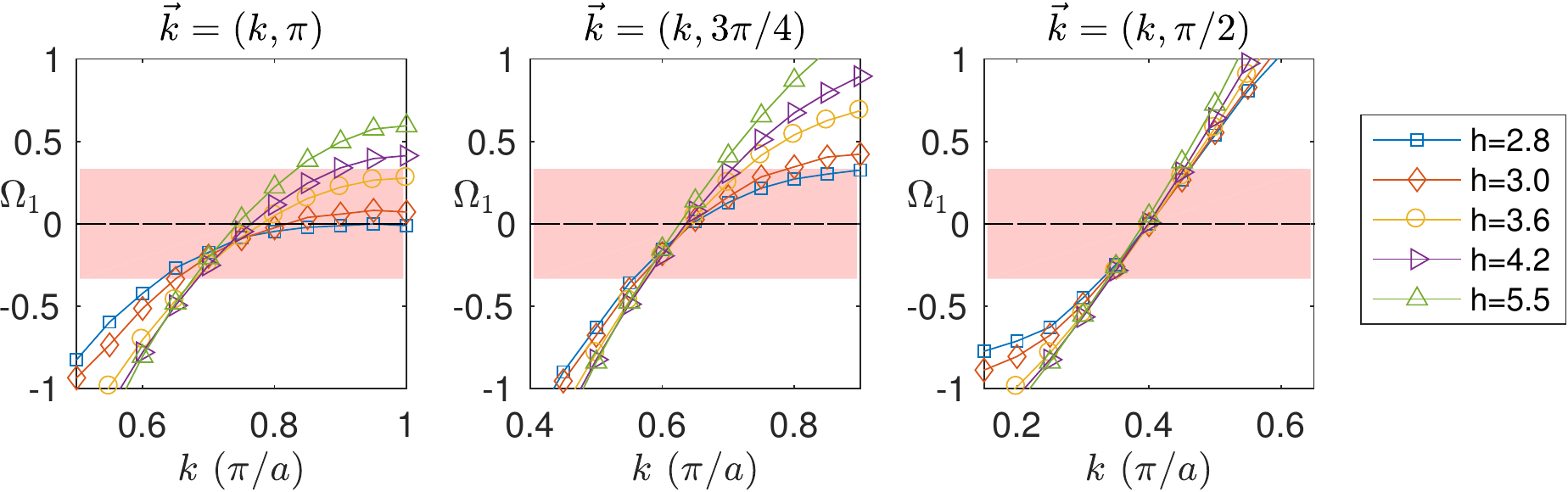}
\caption{The low frequency moment $\Omega_1$ along various cuts through the Fermi surface, showing mass enhancement on approach to the QCP. The QCP is located at $h\approx 2.6$, and for $h\leq h_c$ (upper row), a small symmetry breaking field has been applied to orient the nematic order so that hopping in the $x$ direction is enhanced (this also cuts off fluctuation effects to some extent near $h_c$). The cut through $k_y=\pi/2$ passes near the cold spot, and has little mass enhancement on approach to the QCP. The shaded range is within $2T$ of the Fermi level, roughly where $\Omega_1$ should faithfully measure the quasiparticle dispersion.}
\label{fig:dispersion}
\end{figure}

In Fig. \ref{fig:dispersion} we exhibit the momentum dependence of $\Omega_1$ at fixed temperature for a variety of values of the tuning parameter $h$. Except for near the cold spots, $\Omega_1$ tends to flatten near the Fermi level as $h_c$ is approached. Similar results are found for smaller values of the coupling constant and the fermion density.

\section{Fermion spectral function}
\begin{figure}[h]
    \includegraphics[width=0.9\columnwidth]{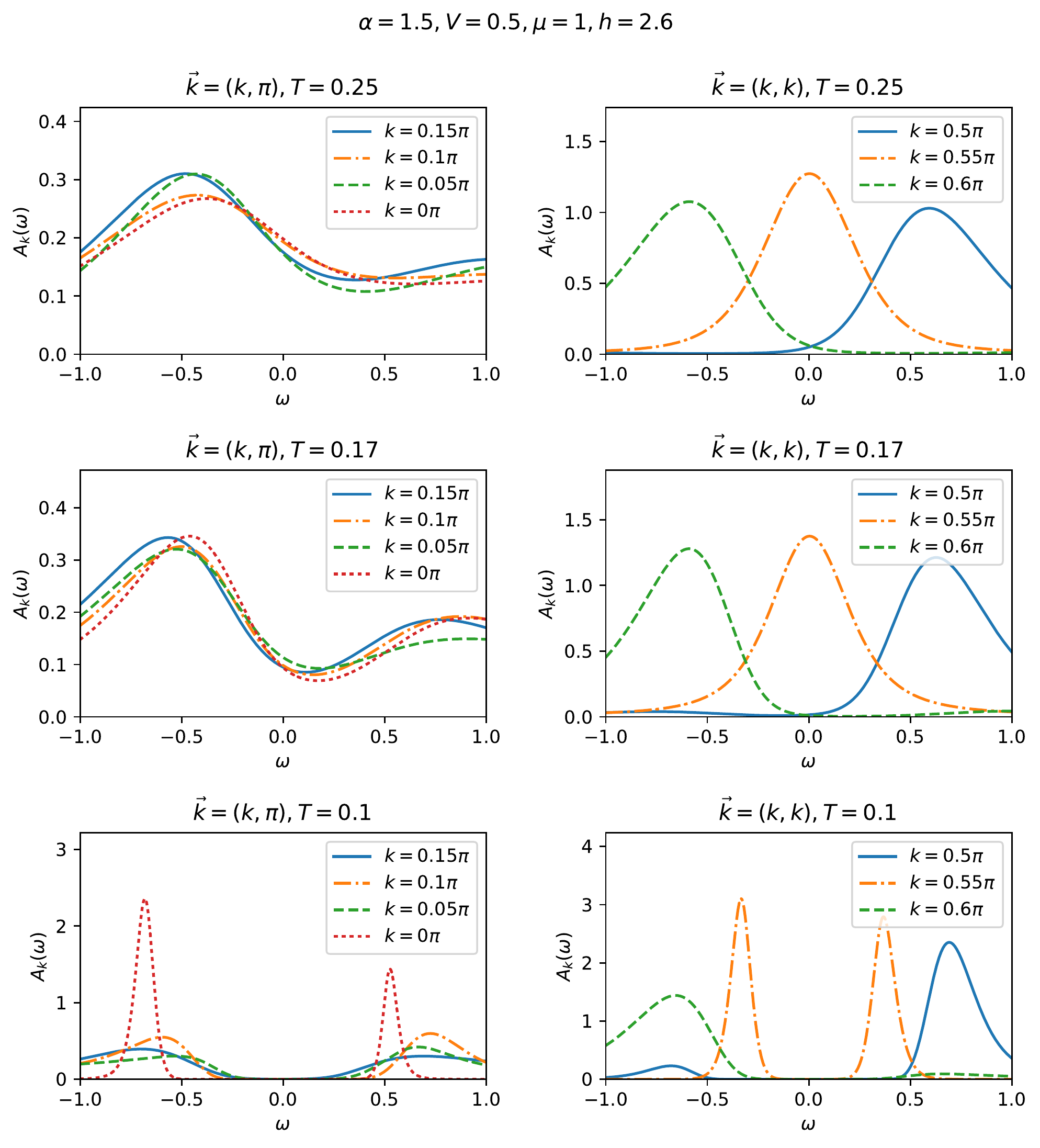}
    \caption{The fermionic spectral function as obtained by the maxEnt method. Different curves represent different momenta close to the Fermi surface. Shown here for $h\approx h_c$}
    \label{fig:maxent_spectral_critical}
\end{figure}

\begin{figure}[h]
    \includegraphics[width=0.9\columnwidth]{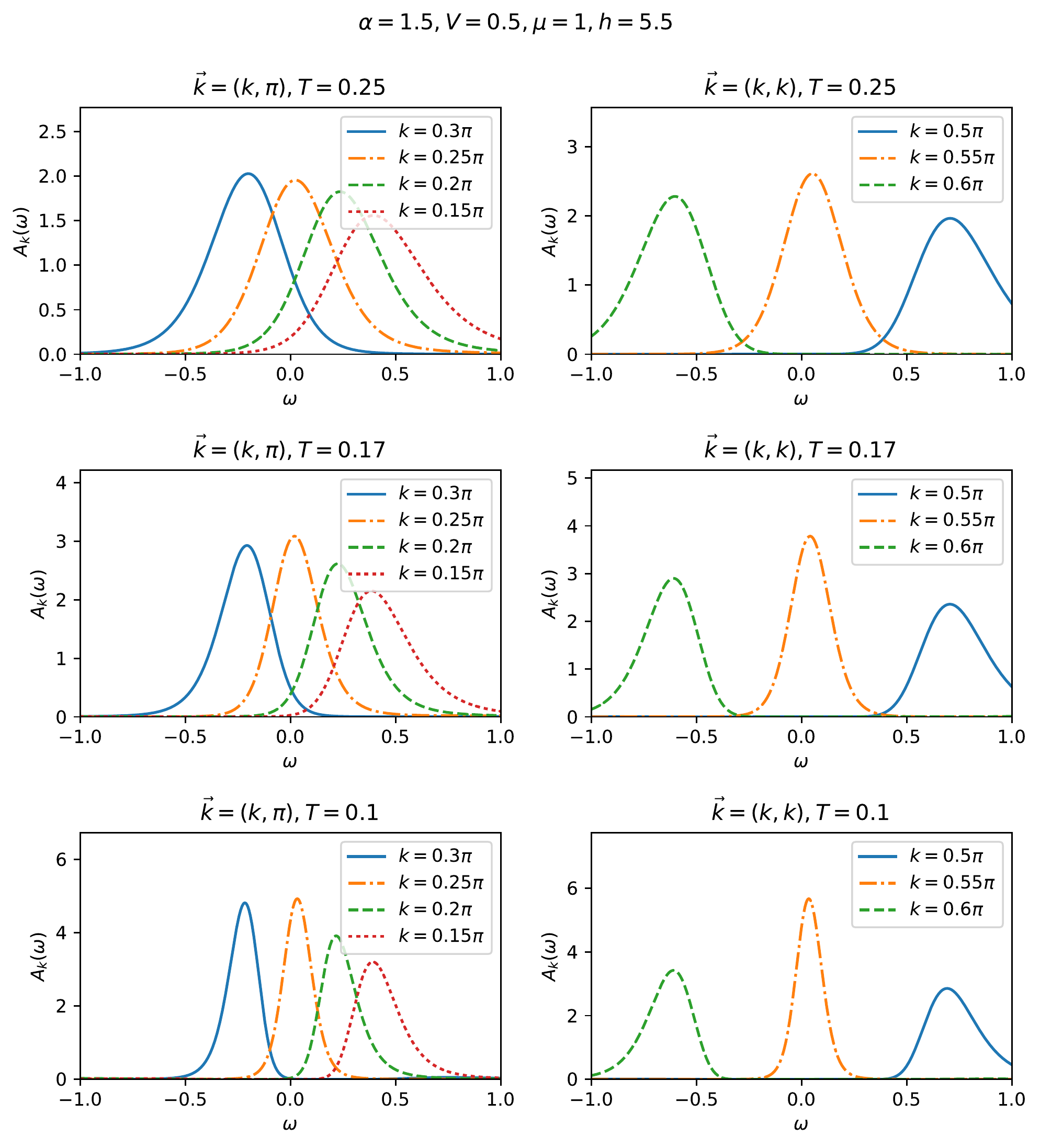}
    \caption{The fermionic spectral function as obtained by the maxEnt method. Different curves represent different momenta close to the Fermi surface. Shown here for $h> h_c$}
    \label{fig:maxent_spectral_disordered}
\end{figure}

In this section we construct the fermionic spectral function $A_{\mathbf{k}}(\omega)$ by the maximum entropy method.
 Close to criticality, [Fig.\ref{fig:maxent_spectral_critical}], the spectral function away from the cold spots shows very broad features, without a well-defined dispersing peak. Below the superconducting $T_c$, a gap opens around $\omega=0$. At the cold spots or away from criticality [Fig.\ref{fig:maxent_spectral_disordered}], we find well-defined peaks, showing a BCS-like transition into the superconducting phase.
Although the maxEnt results are visually appealing and agree with the direct analysis of imaginary-time data shown in the main text, a word of caution is in order. The maxEnt method favors spectral functions which are as smooth and featureless as possible, while still in agreement with the data. Thus, it is not reliable for extracting spectral features with typical frequencies much smaller than the temperature.
\section{Behavior at lower densities}
\begin{figure}[h]
\includegraphics[width=0.45\columnwidth]{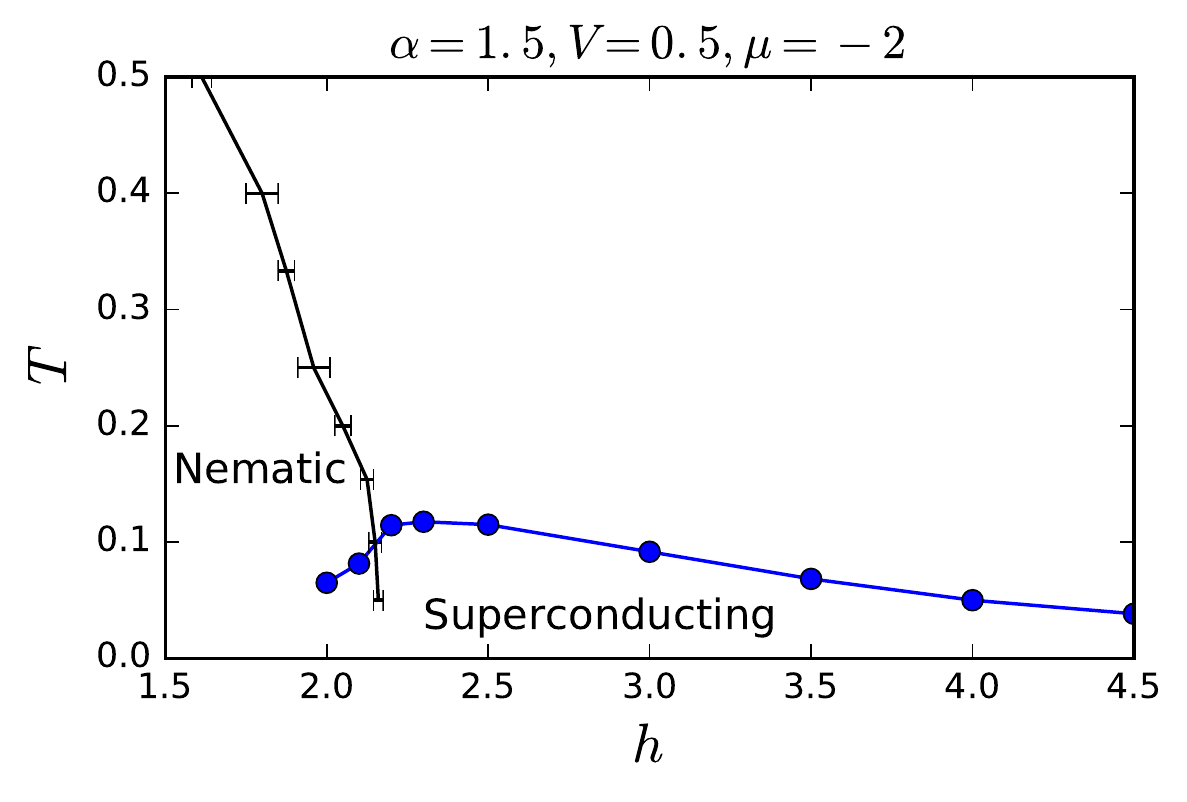} \includegraphics[width=0.45\columnwidth]{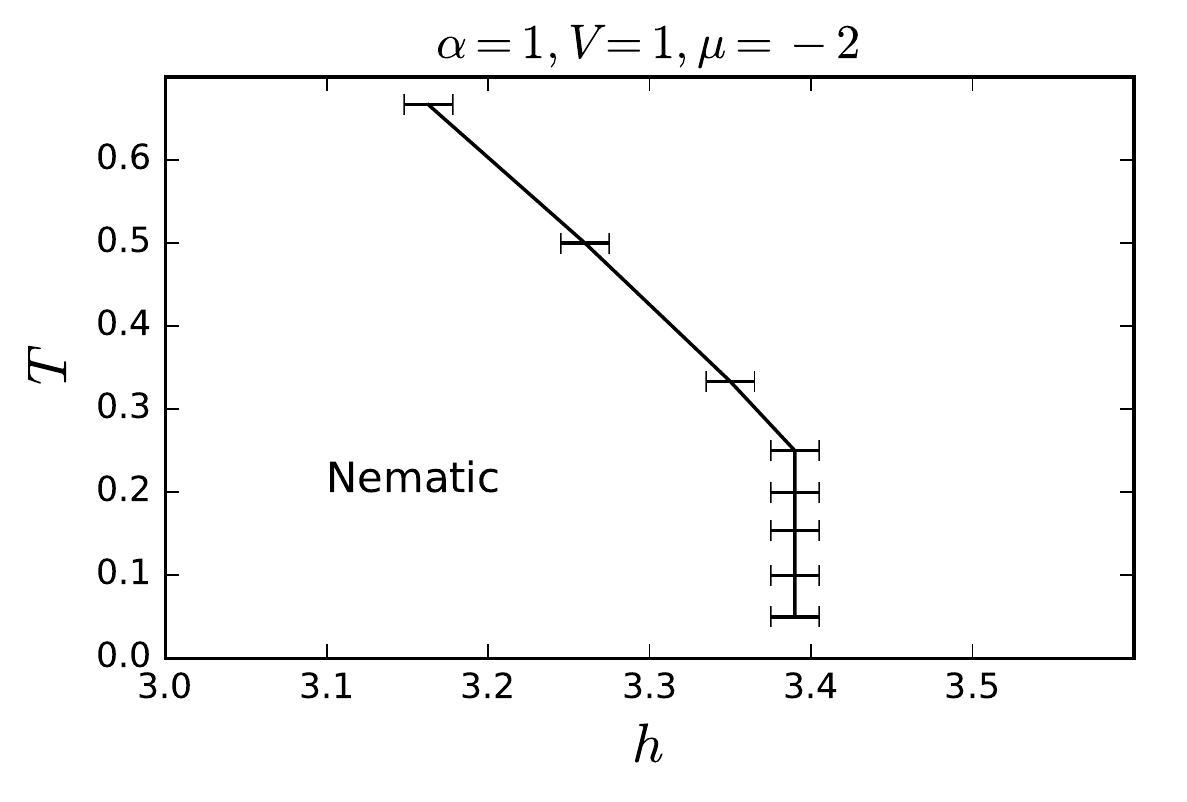}
\caption{Phase diagrams for two values of the coupling constants.}
\label{fig:mu2_phase_diagrams}
\end{figure}
In this section we present results for somewhat smaller densities than those presented in the main text. We focus on the two sets of couplings $\alpha$ and $V$ as in the main text, with the chemical potential set to $\mu=-2$. In figure \ref{fig:mu2_phase_diagrams} we show the phase diagrams. For both couplings, the finite temperature nematic phase boundary $T_\mathrm{nematic}(h)$ is linear at high temperature, and undergoes a sharp change of slope at lower temperatures. At the weaker coupling, $\alpha=1$, a multi-peak structure is seen in the distribution of several thermodynamic quantities, such as the density and the nematic order parameter (not shown), suggesting a weakly first order transition at low temperatures.
Whereas at the stronger coupling, $\alpha=1.5$ and $V=0.5$, we find a high $T_c$ superconducting dome with maximal $T_c \approx 0.12$, there is no evidence of superconductivity for the weaker coupling, $\alpha=1$ and $V=1$, down to temperatures $T = 0.033$.

\begin{figure}[h]
\includegraphics[width=0.8\columnwidth]{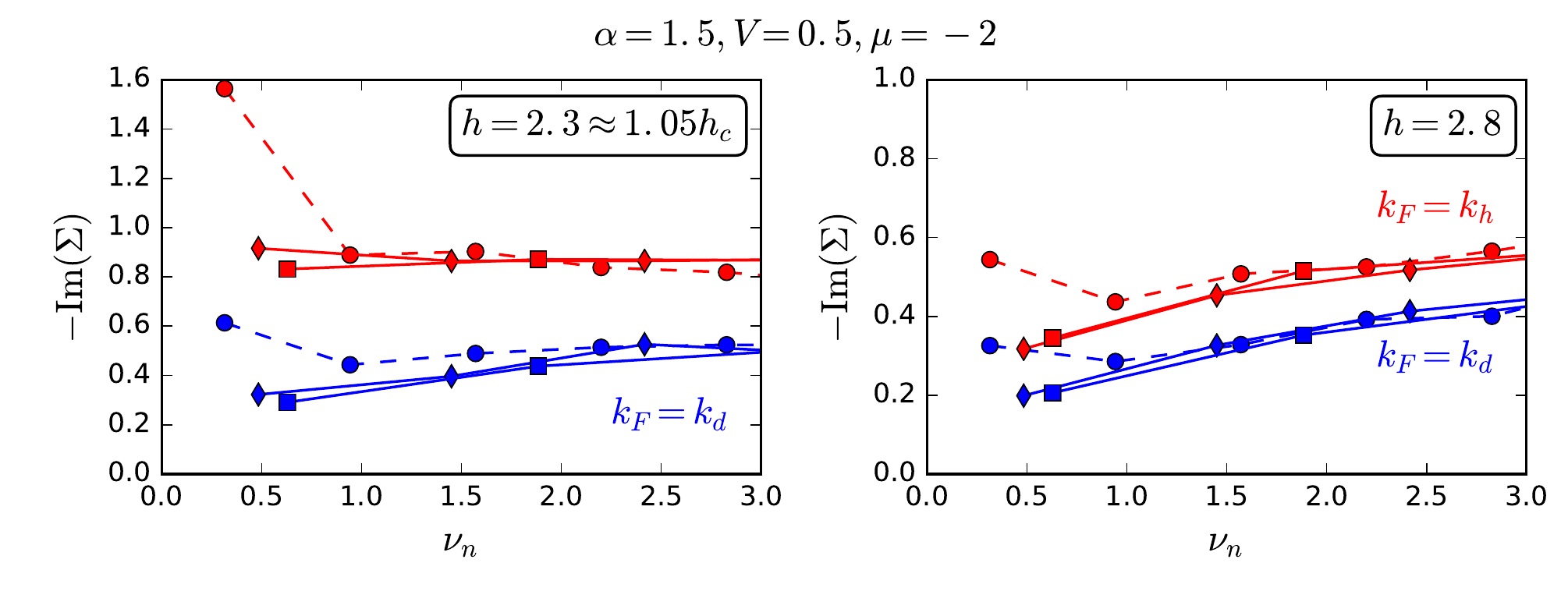}
\includegraphics[width=0.8\columnwidth]{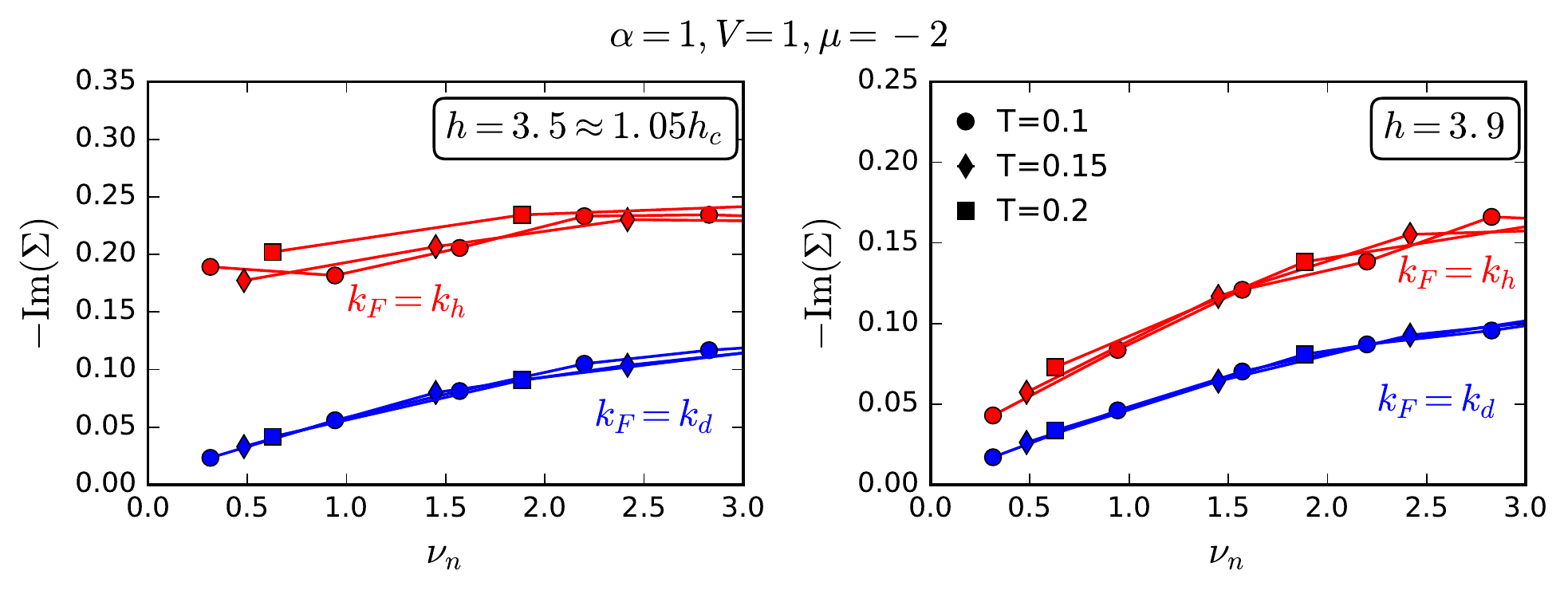}
\caption{The imaginary part of the fermion self-energy, for various temperatures and with the nominal Fermi momenta $k_d$ and $k_h$ along the $(0,0) - (\pi,\pi)$ and $(0,\pi)-(\pi,\pi)$ directions, respectively. Data are shown for a $14 \times 14$ system both near $h_c$ (left column) and somewhat in the symmetric phase (right column). In the upper panels, data points below $T_c$ are connected by dashed lines.}
\label{fig:mu2_sigma}
\end{figure}

The imaginary part of the single fermion self-energy, shown in Fig. \ref{fig:mu2_sigma}, shares similar characteristics to the larger-density data. A substantial ``nodal-antinodal'' dichotomy is seen, and the self energy close to $h=h_c$ seems to approach a constant as $\nu_n\rightarrow 0$. At the stronger coupling, $\alpha=1.5$, the characteristic upturn of the self energy at low frequencies is seen below the superconducting $T_c$.

\begin{figure}[hhh]
\includegraphics[width=0.8\columnwidth]{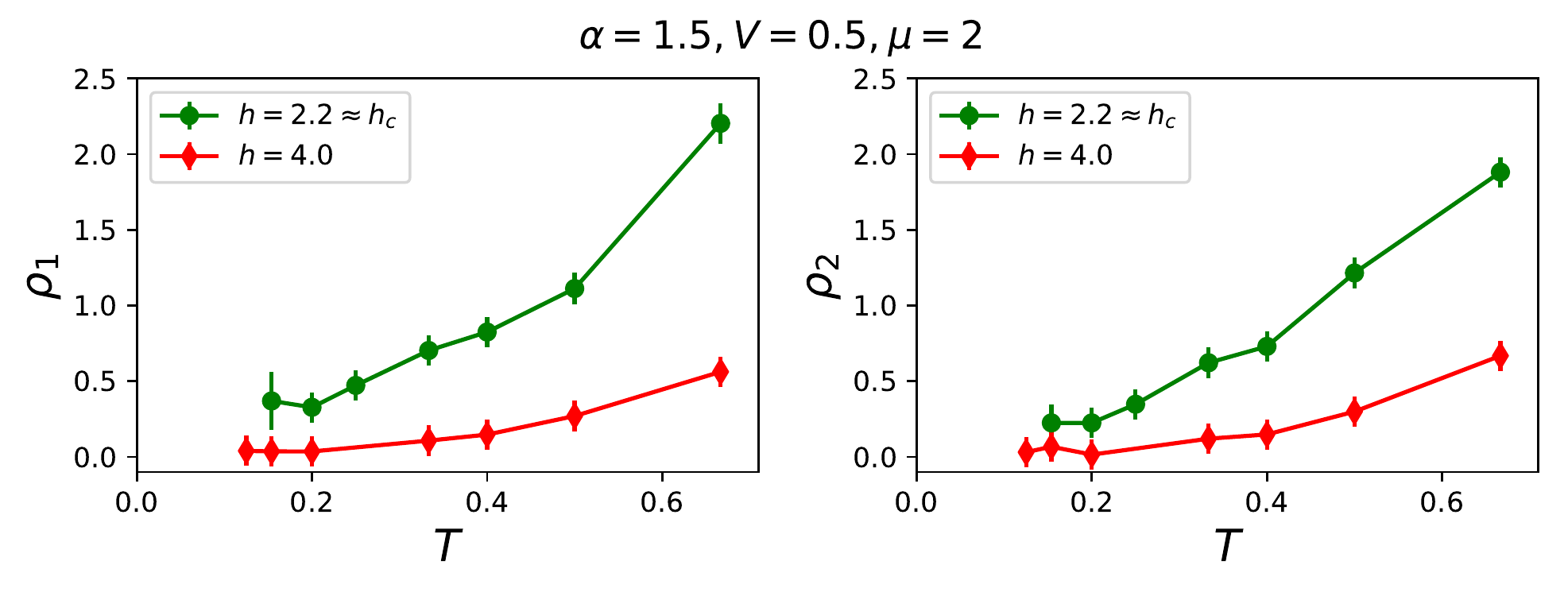}
\includegraphics[width=0.8\columnwidth]{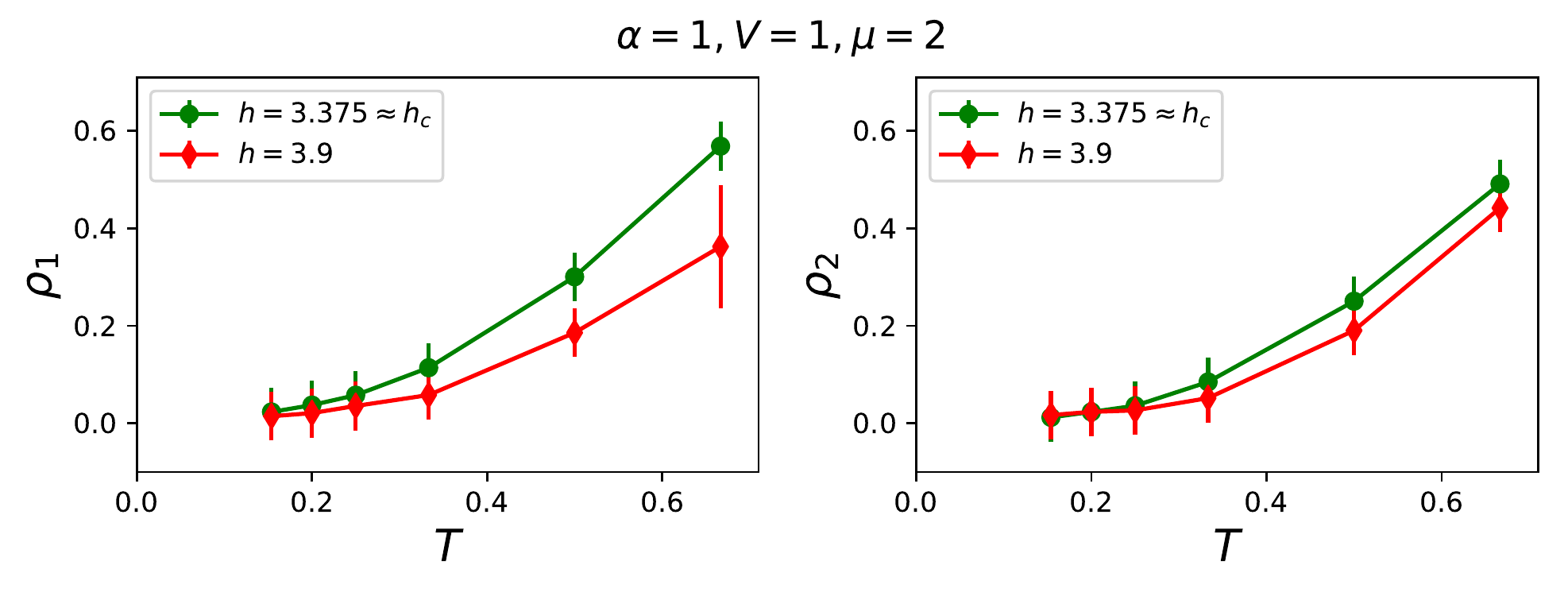}
\caption{The temperature dependence of the resistivity.}
\label{fig:mu2_rho}
\end{figure}

The estimates for the resistivity, $\rho_{1}$ and $\rho_{2}$, are shown in Fig.~\ref{fig:mu2_rho}. As in the higher densities shown in the main text, the resistivity proxy $\rho_{2}$ (right column) agrees qualitatively with the results of the two- component fit  $\rho_{1}$ (left column). The optical conductivity (not shown) contains a Drude-like peak. The magnitude of the DC resistivity is of order of the quantum of resistance $\hbar/e^2$. However, unlike for higher densities, we do not find linear-in $T$ resistivity over a range of temperatures close to the nematic quantum phase transition.

A possible cause for the qualitative difference in the resistivity between the lower and higher density systems  is the smaller size of the Fermi surface at these densities, which might lead to a suppression of certain umklapp processes at low temperatures. For our Fermi surface, an umklapp scattering process involving two fermions near the Fermi point along the line $(0,\pi)-(\pi,\pi)$, $\vec{k}_h$, and symmetry-related points requires $k_h\ge \frac{\pi}{2}$. Similarly, a process involving two fermions near the Fermi points along the diagonal, $\vec{k}_d$, and symmetry-related points requires $k_d \ge \frac{\pi}{\sqrt{2}}$. The magnitudes of the Fermi momenta $k_d$ and $k_h$, measured with respect to $(\pi,\pi)$, are shown in Table \ref{kf_table}, and are found to be close to the aforementioned limiting values.

\begin{table}[h]
\begin{tabular}{|c|c|c|}
\hline
 & $\frac{2k_{h}}{\pi}$ & $\frac{2k_{d}}{\sqrt{2}\pi}$\tabularnewline
\hline
$\alpha=1$,    $V=1$, $\mu=-2$  & $1.16\pm0.08$ & $0.72\pm0.08$\tabularnewline
\hline
$\alpha=1.5$, $V=0.5$, $\mu=-2$ & $1.28\pm0.08$ & $0.80\pm0.08$\tabularnewline
\hline
\end{tabular}\caption{Magnitudes of Fermi momenta along high symmetry directions.}
\label{kf_table}
\end{table}

\end{document}